\documentstyle[preprint,aps,floats,epsf]{revtex}

\begin{document}

%------------------------------------
% New Commands
%------------------------------------
\newcommand{\s}{\rm s}
\newcommand{\id}{\rm ID}
\newcommand{\cm}{\rm cm}
\newcommand{\km}{\rm km}
\newcommand{\gev}{\rm GeV}
\newcommand{\tev}{\rm TeV}
\newcommand{\susy}{\rm SUSY}

\tightenlines %% LANL requests single line spacing to save paper.

%=======================================================================
% TITLE PAGE
%=======================================================================

%-----------------------------------
% Preprint Number
%-----------------------------------
\preprint{\font\fortssbx=cmssbx10 scaled \magstep2
\hbox to \hsize{
\hfill$\vcenter{\hbox{\bf MADPH-03-1345}
                \hbox{\bf hep-ph/0309036}
                \hbox{September 2003}}$ }
}

%-----------------------------------
% Title
%-----------------------------------
\title{\vspace{.5in}
Direct and Indirect Detection of Neutralino Dark Matter \\ In Selected Supersymmetry Breaking Scenarios}

%-----------------------------------
% Authors
%-----------------------------------
\author{
Dan Hooper$^{1,2}$ and Lian-Tao Wang$^2$}

%-----------------------------------
% Address
%-----------------------------------
\address{
$^1$ Denys Wilkinson Laboratory, Astrophysics Department, OX1 3RH Oxford, England, UK \\
$^2$ Department of Physics, University of Wisconsin,
Madison, WI 53706, USA
}

\maketitle

\bigskip

%-----------------------------------
% Abstract
%-----------------------------------
\begin{abstract}

Various methods of searching for supersymmetric dark matter are
sensitive to WIMPs with different properties. One consequence of this is that the phenomenology of dark matter detection can vary
dramatically in different supersymmetric breaking scenarios.   

In this paper, we consider the sensitivities to supersymmetric dark
matter of different detection methods and techniques in a wide variety of supersymmetric breaking scenarios. We discuss the ability of various astrophysical experiments, such as direct experiments, gamma-rays satellites, neutrino
telescopes and positron and anti-proton cosmic ray experiments, to test various supersymmetry breaking scenarios. We also discuss what information can be revealed about supersymmetry breaking by combining results from complementary experiments. We place an emphasis on the differences between various experimental techniques.

\end{abstract}

%

%=======================================================================
%   BEGIN MAIN TEXT
%=======================================================================
\newpage

%-----------------------------------------------------------------------
% 1. Introduction
%-----------------------------------------------------------------------
\section{Introduction}

\vspace{0.2in}

An enormous body of evidence has accumulated in the favor of cold
dark matter. This body of evidence includes observations of galactic
clusters and large scale structure \cite{structure}, supernovae
\cite{supernovae} and the cosmic microwave background (CMB) anisotropies
\cite{cmb,wmap}. Recently, WMAP has provided the most detailed
information on the CMB to date, quoting a total matter density of
$\Omega_{\rm{m}} h^2 = 0.135^{+0.008}_{-0.009}$
\cite{wmap}. Furthermore, data from WMAP and other prior experiments
indicate a considerably smaller quantity of baryonic matter
\cite{wmap,nonbaryonic}. At the 2-$\sigma$ confidence level, the
density of non-baryonic, and cold, dark matter is now known to be
$\Omega_{\rm{CDM}} h^2 = 0.113^{+0.016}_{-0.018}$ \cite{wmap}. 

A compelling candidate for the cold dark matter is naturally provided
by supersymmetry \cite{susylsp}. In supersymmetric models which
conserve R-parity \cite{rparity}, the lightest supersymmetric particle
(LSP), is stable. Furthermore, in many supersymmetric models, the LSP
is the lightest neutralino, a mixture of the superpartners of the
photon, Z and neutral higgs bosons, and is electrically
neutral, colorless and a viable dark matter candidate.   
If such an particle were in equilibrium with photons in the early universe,
as the temperature decreased, a freeze-out would occur leaving a thermal
relic density. The temperature at which this occurs, and
the density which remains, depends on the annihilation cross section
and mass of the lightest neutralino. It is natural for supersymmetry to provide
a dark matter candidate with a present abundance similar to those
favored by the WMAP experiment.  
				        	
Many methods have been proposed to search for evidence of neutralino
dark matter. In addition to accelerator searches for supersymmetry
\cite{collider}, direct and indirect dark matter detection experiments
have been performed. Direct dark matter searches attempt to measure
the recoil of dark matter particles scattering elastically off of the
detector material \cite{direarly,direct}. Indirect dark matter
searches hope to observe the products of dark matter annihilation
including gamma-rays
\cite{indirectgamma,continuum,loop,brenda,bergstrom}, neutrinos
\cite{indirectneutrino}, positrons \cite{positrons,positronswang,baltzpos} and
anti-protons \cite{antiprotons}. 
 
Supersymmetry with low scale soft
breaking has been established as an excellent candidate for physics beyond
the standard model. At low energy, a usual practice in studying
supersymmetry phenomenology is to parameterize the supersymmetry
breaking by putting in explicit soft supersymmetry breaking terms in
the effective Lagrangian. In principle, all of those terms should be
derived from some supersymmetry breaking and mediation mechanism which
is embedded in the fundamental theory. Therefore, supersymmetry breaking
mechanisms provide the crucial link between the fundamental 
theory and the low energy parameters. Knowledge of supersymmetry
breaking will provide us with important information about the physics at
the high energy scale. We would like to extract as much information as
possible about how supersymmetry is broken from our low energy
experiments. Needless to say, the identity and properties of the LSP would carry
invaluable information about the underlying supersymmetry breaking
mechanism. In this paper, we attempt to evaluate the prospects for direct and
indirect detection of neutralino dark matter in a variety of 
supersymmetry breaking scenarios. We find
distinctive differences in 
the dark matter phenomenology for various breaking mechanisms.
Information from cold dark matter
experiments may be very useful in providing hints about which
supersymmetry breaking scenario is manifest in nature. 

This paper is organized as following: First we briefly review various
direct and indirect detection methods for cold dark matter. We
then review the relevent information regarding the different supersymmetry
breaking scenarios that we study in this paper. Finally, we present our
results and comparison of different supersymmetry breaking scenarios.

\section{Direct Detection}

\vspace{0.2in}

If the dark matter halo of our galaxy is made up of TeV-scale
particles, then millions of such particles travel through each square
meter each second in our galaxy.  Despite this large rate, dark matter
particles likely have very small cross sections, making
observation of such particles very difficult.   

Proposed as early as twenty years ago \cite{direarly}, numerous
experiments have been developed in an effort to directly detect dark
matter particles. In this paper, we will compare theoretical
predictions to recent experimental results and to the reach of future
efforts. 

The ability of a direct detection experiment to observe a WIMP depends
on that particle's elastic scattering cross section with nucleons in
the detector's target medium. This quantity depends strongly on the
composition of the lightest neutralino (gaugino verses higgsino) as
well as the masses of the squarks mediating scattering
processes. 

Limits or
sensitivities are most frequently shown as a elastic scattering cross
section verses WIMP mass.  Since the lightest neutralino is a Majorana
fermion, it 
only has axial-vector and scalar interactions with nuclear matter. For
heavy nuclei, typically $A>20$, the scalar (or spin independent)
interaction dominates. Therefore, we focus on the limits and reaches
for scalar interaction cross sections.  For                                                                                                       
each model, we assume the measured local dark matter                                                      
density and velocity distribution \cite{local}. 

There are a large number of direct detection experiments which have
been developed or are planned.  Currently, the strongest sensitivities
have been attained by the CDMS \cite{cdmsnow}, ZEPLIN-I
\cite{zeplin1}, and Edelweiss \cite{edelweiss}
experiments. Additionally, the DAMA collaboration has claimed an
observation of an annual modulation in their rate consistent with a
$\sim 50\,$ GeV WIMP with $\sigma_{\chi N}\sim 10^{-5}\,$ pb
\cite{dama}. The DAMA result has been highly controversial and the
entire region claimed by the experiment is now excluded by other
experiments. 

The next round of direct experiments include CREST-II \cite{crest},
CDMS (Soudan site) \cite{cdms}, Edelweiss II \cite{edelweiss2} and
ZEPLIN-II, III, and IV \cite{zeplin4}.  These experiments will be
capable of studying WIMPs with considerably lower cross sections in
the relatively near future. Later, experiments such as XENON
\cite{xenon} or GENIUS \cite{genius} may probe even further.

\section{Indirect Detection}

\vspace{0.2in}

Several different methods have been pursued to search for cold dark matter particles indirectly. In these approaches, experiments are designed to look for annihilation products from WIMP annihilations, such as neutrinos from the Sun or Earth, gamma-rays from the center of the galaxy, and positrons or anti-protons from the galactic halo.

\subsection{Gamma-Rays From The Galactic Center}

\vspace{0.2in}

In regions where the dark matter density is very large, the annihilation rate may become large enough to provide observable
fluxes of high energy gamma-rays (and also neutrinos). The spectral \cite{gammaspecold} and angular
\cite{gammaanglold} features of such a source have been studied for decades. The center of our
galaxy may be such a region. Satellite-based
experiments, such as EGRET and GLAST, are capable of observing gamma-rays at the energies suitable for this purpose.

The rate at which neutralinos annihilate near the galactic center depends strongly on the halo profile considered. At this time, the distribution of galactic dark matter, especially near the galactic center, is not well known.  The halo models commonly discussed in this debate include distributions with low density cores, high density cusps or even higher density spikes.  

N-body simulations appear to predict models with cuspy distributions,
such as the Navarro, Frenk and White (NFW) or Moore, {\it et. al.}
profiles \cite{cusp}. In such models, the dark matter density
increases as, $\rho \propto 1/r^{\gamma}$, near the central
region. $\gamma$ is 1.0 in the NFW profile and 1.5 in the Moore, {\it et. al.}
profile. Arguments have been made, however, that observations of
galactic rotation curves favor
flat density core profiles \cite{observations}, although others claim
that these observations do not preclude the presence of a cusp at the
center of our galaxy \cite{however}. In the case of a core profile, no observable gamma-ray signal would be produced in dark matter annihilations near the galactic
center region. 

Models with substantial central density spikes have been discussed a
great deal recently \cite{spike}. It has been argued that density
spikes are generated naturally as a result of adiabatic accretion of
matter into the central galactic black hole. If density spikes form in
the center of galaxies, it would be possible for dark matter
annihilation rates to occur near the center of our galaxy which were
considerably larger than those predicted for cuspy or other halo
profiles.  
 
There are two signatures for the detection of gamma-rays from the
galactic center which we will consider in this paper. The first is
line emission produced from the processes $\chi^0 \chi^0 \rightarrow
\gamma \gamma$ and $\chi^0 \chi^0 \rightarrow \gamma Z$. The second is
continuum emission from all gamma-ray producing processes, dominated
by $\pi^0$ decay created in the fragmentation of quarks
\cite{continuum}. Although the continuum emission lacks the
distinctive spectral features of the line emission, the number of
photons produced is considerably greater. Since there is no direct
coupling between the neutralino and the photon, all line producing processes must be
mediated by some charged particle loop (such as SM fermions, charginos,
W-bosons, etc.). Therefore, all line emission processes
are necessarily loop suppressed.  For all of the diagrams contributing to line
emission at the one loop level, see \cite{loop}. $\gamma \gamma$ and $\gamma Z$ line emission will give rise to very
distinct feature in the spectrum (since the end products will have a very narrow energy distribution). The position of these lines can also be used to measure the mass of the annihilating WIMP. This is probably the only way to accurately measure the WIMP's mass outside of a collider.

\subsection{Neutrinos From The Sun}

\vspace{0.2in}

In addition to the galactic center region, high densities of dark
matter may accumulate in the center of less distant objects such as
the Sun and Earth. In these deep gravitational wells, the annihilation
rate of WIMPs can be large, resulting in observable fluxes of
annihilation products, such as neutrinos \cite{oldneutrinos}. High-energy neutrino
telescopes, such as AMANDA \cite{amanda}, ANTARES \cite{antares} or
next generation IceCube \cite{icecube}, are designed, in part, for
this task. For a review of high-energy neutrino astronomy, see
Ref.~\cite{neutrinoreview}. 

The flux of neutrinos from neutralino annihilations in the Sun or Earth
is a function of the annihilation cross section of the WIMP, the
WIMP-nucleon elastic scattering cross section and the ratios of the
various WIMP annihilation modes. For most SUSY models, the
annihilation rate reaches equilibrium with the capture rate in the
Sun, and the dependence on the annihilation cross section is
removed. Typically the flux of neutrinos, and the associated event
rate, from WIMP annihilation in the Earth is much far smaller than
from the Sun, so we will consider only the rate from the Sun in this
paper. 

Neutrinos can be produced in neutralino annihilation by several
processes including $\chi^0 \chi^0 \rightarrow t \bar{t}$, $\chi^0
\chi^0 \rightarrow b \bar{b}$, $\chi^0 \chi^0 \rightarrow c \bar{c}$,
$\chi^0 \chi^0 \rightarrow ZZ$, $\chi^0 \chi^0 \rightarrow W^+ W^-$
and $\chi^0 \chi^0 \rightarrow \tau^+ \tau^-$
\cite{jungman}. Neutrinos are produced directly in the decays of
$\tau^{\pm}$'s, $c$ and $b$ quarks and gauge bosons, and indirectly
through the decays of the $b$ quarks and $W^{\pm}$ bosons created in
top quark decays. The process(es) which dominates depends on the mass
and composition of the lightest neutralino. In the fragmentation
process, particles can lose substantial amounts of their
annihilation energy before they decay and produce a
neutrino. Furthermore, high-energy neutrinos can interact as they
propagate through the Sun, thus degrading their flux
\cite{jungman,crotty}.

\subsection{Positrons and Anti-Protons From The Galactic Halo}

\vspace{0.2in}
 
Annihilating halo neutralinos can also generate positrons and
anti-protons. Unlike gamma-rays or neutrinos, charged cosmic rays do
not point at their sources due to galactic magnetic fields. By measuring the flux of positrons and
anti-protons at Earth, it may be possible to observe signatures of
annihilating neutralinos in nearby dark matter clumps. 

In 1994 and 1995, the High Energy Antimatter Telescope (HEAT) observed
an excess of cosmic positrons, peaking around $\sim10\,$ GeV
\cite{heat1995}. This result was confirmed by another HEAT flight in
2000 \cite{heat2000}. Although it is uncertain whether this excess is
the result of dark matter annihilation, it has been shown that halo positrons
could be a signature of such a process \cite{positrons,positronswang,baltzpos}. 

Using a smooth halo profile, we calculate the positron flux on Earth
following the methods of \cite{baltzpos}. This includes not only the
calculation of the injection rate of positions from neutralino
annihilation, but also the propagation effects such as synchrotron and
inverse Compton energy losses, which are very important. We find that for a smooth halo distribution, the positron flux predicted is probably insufficient to
explain the HEAT data. Any deviations from a smooth halo profile,
however, enhance the positron rate, especially if clumps occur
relatively nearby. We parameterize this effect with a single variable,
the positron boost factor, which is the ratio of the positron flux at Earth for
a given halo profile to the positron flux for a smooth profile.  

The anti-proton flux can also be calculated
\cite{antiprotons}. Anti-protons can propagate much greater distances
than positrons, however, often traveling up to ten or more
kiloparsecs. This requires a separate boost factor parameter for
anti-protons. The most relevant measurement of the anti-proton flux comes from the 1995 and 1997 BESS (Balloon Borne
Experiment with
Superconducting Spectrometer) data \cite{bess}. The anti-proton flux
measured by BESS is $1.27^{+0.37}_{-0.32} \times 10^{-6} \rm{cm}^{-2}
\rm{s}^{-1} \rm{sr}^{-1} \rm{GeV}^{-1}$ in the range of 400 to 560
MeV. 

\section{Supersymmetry Breaking}    

\vspace{0.2in}

The general framework to study the phenomenology of softly broken
low energy supersymmetry is the Minimal Supersymmetric Standard Model
(MSSM). The matter spectrum of the MSSM can be obtained basically by
assigning a superpartner to each of the Standard Model fields. The
only exception is that in the MSSM, we require two Higgs
doublets.  

\noindent The MSSM superpotential is given by
\begin{equation}
\label{superpot}
W=\epsilon_{ab}[-\hat{H}_u^a\hat{Q}_i^bY_u^{ij}\hat{U}^c_j
+\hat{H}_d^a\hat{Q}_i^bY_d^{ij}\hat{D}^c_j
+\hat{H}_d^a\hat{L}_i^bY_e^{ij}\hat{E}^c_j-\mu\hat{H}_d^a\hat{H}_u^b],
\end{equation}
in which $\epsilon_{ab}=-\epsilon_{ba}$ and $\epsilon_{12}=1$, and
the superfields are defined in the standard way (suppressing gauge 
indices):
\begin{eqnarray}
\label{superfielddef1}
\hat{Q}_i&=&(\tilde{Q}_{L_i}, Q_{L_i})\nonumber\\
\hat{U}^c_i&=&(\tilde{U}^c_{L_i}, U^c_{L_i})\nonumber\\
\hat{D}^c_i&=&(\tilde{D}^c_{L_i}, D^c_{L_i})\nonumber\\   
\hat{L}_i&=&(\tilde{E}_{L_i}, E_{L_i})\nonumber\\
\hat{E}^c_i&=&(\tilde{E}^c_{L_i}, E^c_{L_i})\nonumber\\
\hat{H_u}&=&( H_u, \tilde{H}_u)\nonumber\\
\hat{H_d}&=&( H_d, \tilde{H}_d),
\end{eqnarray}  
with $i,j=1\ldots 3$ labeling family indices.
The soft breaking Lagrangian, ${\mathcal{L}}_{soft}$, takes the form
(dropping
``helicity" indices):
% want $-$ sign here to be in the convention for RGEs.
\begin{eqnarray}
\label{lsoftexpr}
-{\mathcal{L}}_{soft} &=&\frac{1}{2}\left[M_3
\tilde{g}\tilde{g}+M_2\tilde{W^a}\tilde{W^a}+M_1\tilde{B}\tilde{B}\right
]\nonumber\\
&+&\epsilon_{ab}[-b H^a_dH^b_u-
H^a_u\tilde{Q}^b_i\tilde{A}_{u_{ij}}\tilde{U}^c_j
+H^a_d\tilde{Q}^b_i\tilde{A}_{d_{ij}}\tilde{D}^c_j
+H^a_d\tilde{L}^b_i\tilde{A}_{e_{ij}}\tilde{E}^c_j +h.c.]\nonumber\\
&+&m_{H_{d}}^{2}|H_{d}|^2+m_{H_{u}}^{2}|H_{u}|^2
+\tilde{Q}^a_i{m^2_Q}_{ij}\tilde{Q}_j^{a*} \nonumber\\
&+& \tilde{L}^a_i{m^2_L}_{ij}\tilde{L}_j^{a*}
+\tilde{U}^{c*}_i{m^2_U}_{ij}\tilde{U}^c_j
+\tilde{D}^{c*}_i{m^2_D}_{ij}\tilde{D}^c_j
+\tilde{E}^{c*}_i{m^2_E}_{ij}\tilde{E}^c_j.
\end{eqnarray}
The $m^2$'s, usually called soft masses squared, are in general
$3 \times 3$ Hermitian matrices. $\tilde{A}$ are usually called
trilinears which are $3 \times 3$ general complex matrices. In many
supersymmetry breaking scenarios where $\tilde{A}$ is universal, such as mSUGRA
(which we will review later), it is common to write $\tilde{A}=Y \cdot
A$. $M_i$'s are the mass terms for the superpartners of the Standard
model gauge bosons, gauginos. 

Notice that the MSSM by itself is not a model of supersymmetry
breaking. It is a parameterization of supersymmetry breaking by
including general soft breaking terms explicitly. 
As a result, the MSSM has a large parameter space containing 124
parameters. Most of these are soft SUSY breaking parameters related to
the masses and mixings of the superpartners, which are not
measured. It is impossible to scan the full parameter space of the
general MSSM for phenomenological purposes. On the other hand, all of
the soft parameters are presumably generated by some underlying
supersymmetry breaking and mediation mechanism. Usually, such a
mechanism will provide relations between the MSSM soft parameters and 
reduce the number of parameters significantly. The properties make it
possible, after the detection of low energy
supersymmetry, to distinguish different supersymmetry breaking
scenarios based  on experimental data. Since the LSP appears 
naturally in the MSSM as a natural candidate for cold dark matter, it is
interesting to study the prospects for dark matter
observations in different SUSY breaking scenarios. 

Most of the scenarios we consider in this article fall into the
general category of gravity mediated supersymmetry breaking
\cite{gravitymediation}. Gravity is 
a natural candidate to mediate supersymmetry breaking because regardless
of other details of the model, the
gravitational interaction between the hidden and the observed sector
exists. For some other mediation mechanism to dominate, one must
suppress the contribution from gravity. For this reason, many benchmark scenarios have been
constructed and studied in the gravity mediation framework. The main
alternative to gravity mediation is supersymmetry breaking
mediated by gauge interactions \cite{gaugemediation}. In the gauge mediation
scenario, we have approximately the following relation between the soft masses
and the gravitino mass 
\begin{equation}
\frac{m_{\tilde{g}}}{m_{\mathtt{soft}}} \sim
\frac{1}{\alpha_a}\frac{M_S}{M_{Pl}} \ll 1, 
\end{equation}
where $M_S$ is some typical supersymmetry breaking scale. Therefore,
generically, we will have a very light gravitino as the LSP
\cite{gaugemediationdark}. Since we 
focus our attention in the paper on neutralino dark matter,
we will not consider gauge mediated scenarios in detail. 

Before discussing the detailed study, we briefly
review the different scenarios we have considered. Notice that the
patterns of soft parameters resulting from different SUSY breaking
scenarios are given at the input scale, which we take to be the grand
unification scale. The phenomenologically interesting low energy
parameters are obtained through the running of the renormalization
group equations.   

\subsection{mSUGRA}
The general framework to study gravity mediated supersymmetry breaking
is the N=1, D=4 supergravity Lagrangian which is an effective Lagrangian
after integrating out quantum gravity effects at the Planck scale. 
Without  a specific model, general gravity mediation will not give
rise to specific predictions and relations between different
parameters. The most general supergravity Lagragian would contain all
of the terms allowed by gauge symmetry.  However, a minimal benchmark in
this scenario was 
constructed \cite{gravitymediation} and well studied as an example of the 
physics of gravity mediated supersymmetry breaking. The main
ingredients of this scenario (at the supersymmetry breaking scale) are
\begin{enumerate}
\item Flavor diagonal:

The left-left and right-right blocks of sfermion mass matrices are diagonal and
proportional to the unit matrix. The trilinear couplings are also
universal. This type of sfermion mass matrix is primarily motivated by the
phenomenological consideration of satisfying flavor constraints. 

\item Universality:

All of the diagonal soft masses are assumed to have the same value,
$m_0$. The diagonal trilinears are also  assumed to be the same, with
value $A_0$. This is a 
assumption motivated largely by the simplicity of the parameter space. 

\item Universality of the gaugino masses, $M_{1/2}$: 

Gauge unification, in simple GUT scenarios, implies
the unification of the gaugino masses. Therefore, the mSUGRA choice of
gaugino mass universalilty is primarily motivated by gauge coupling
unification. Notice however, the link between these two issues is not always as direct as in the
simplest scenario. Hence, as with the
requirement of gauge unification, we should still treat gaugino mass
unification as an assumption.
%The LSP in this case is main Bino. 
\end{enumerate}

Under those assumptions, the mSUGRA model contains the following set of
parameters
\begin{equation}
m_0, \hspace{0.3cm} A_0, \hspace{0.3cm} M_{1/2}, \hspace{0.3cm} \tan
\beta,  {\mathtt{Sign}}(\mu). 
\end{equation}

In most of the parameter space of the mSUGRA model, the LSP is Bino-like (the superpartner of the hypercharge gauge boson).

We note that despite its simplicity, mSUGRA only represents a very
special corner of the possible model space of supersymmetry
breaking. Most of the universality assumptions may not be present in other scenarios while
still satisfying the motivations of
gauge unification and respecting flavor constraints. 

\subsection{Anomaly Mediated Supersymmetry Breaking}

The authors of Ref.~\cite{anomalymediation} pointed out an important
source of supersymmetry 
breaking mediation in the supergravity Lagrangian due to the so-called
super-conformal anomaly. It produces a elegant solution to the flavor
problem due to the ``UV-insensitivity'' of the soft parameters in the
sense that they can be expressed purely in terms of the low energy
parameters such as the Yukawa couplings and gauge couplings.
Reviewing the details of the origin of 
such a source is outside of the scope of this paper. Instead, we will focus on
the characteristics of the soft spectrum as a result of anomaly
mediation. The soft spectrum of the minimal anomaly mediation is
\cite{anomalymediation,amsb} 

\begin{eqnarray}
M_{\lambda^a} &=& \frac{\beta_{g_a}}{g_a} m_{3/2}, \nonumber \\
m^2_{\tilde{f}} &=& -\frac{1}{4} \left( \frac{\partial \gamma }{\partial
g } \beta_g + \frac{\partial \gamma }{\partial y } \beta_y \right)
m_{3/2}^2 + m_0^2, \nonumber \\
A_y &=& - \frac{\beta_y}{y} m_{3/2},
\end{eqnarray}
where $y$ collectively denotes the Yukawa couplings. The
$\beta$-functions and anomalous dimensions, $\gamma$, are computed in the
supersymmetric limit \cite{anomalymediation,amsb}
and, therefore, are functions of the gauge couplings 
and superpotential parameters.  $m_0$ is some addition parameter
outside of the AMSB scenario, not to be confused with the universal scalar mass in the
SUGRA scenario. It is added by hand mainly to give rise
to a non-tachyonic stau mass. In the minimal AMSB scenario, the scalar
masses squared at the GUT scale are each given by: 

\begin{equation}
m^2_{\tilde{S}} = m_0^2 -\frac{1}{4} \left( \frac{\partial \gamma }{\partial
g } \beta_g + \frac{\partial \gamma }{\partial y } \beta_y \right)
m_{3/2}^2 + \rm{D}\,\,\rm{terms}. \nonumber \\
\end{equation}
However, in non-minimal scenarios, the first term may vary. 

The most obvious feature of the anomaly mediation is the
proportionality of the $\beta$-function to the soft parameters. Since
the origin of the mediation mechanism is the super-conformal anomaly,
or the scale dependence of the parameters of the Lagrangian, it is not
surprising to see this proportionality to the slope of the RGE running of
the parameters. Notice some important features of the anomaly
mediation spectrum:
\begin{enumerate}
\item The main phenomenological implications of this scenario are that
  the gaugino masses have the ratios:  
\begin{equation}
M_1:M_2:M_3=2.8:1:7.1
\end{equation}
leading to the LSP being a neutral wino \cite{amsb,amsbdark}, which is only slightly  
lighter than the charged wino (by a few hundred MeV).
This leads to a long lived lightest chargino with a distinctive collider
signature. Notice that in the AMSB scenario, the gluino is usually quite heavy
due to its large input scale value. This will in turn lead to large
squark masses, since typically $m_{\tilde{q}}^2 \sim c_0
m_{\tilde{q}}^2 (0) + c_3 M_3^2(0)$, where $c_0 \sim O(1)$ and $c_3
\sim 5-7$. 
\item  Unfortunately, the slepton masses squared turn out to be
negative due to the fact that the pure anomaly mediation contribution
to its value is negative. This is clearly unacceptable as it will lead
to a charge 
breaking minima. This is also the reason behind the introduction of
the parameter $m_0$ as it parameterizes additional physics which may solve
this problem. However, introducing such a parameter undermines one of
the main
advantages of the AMSB scenario. Since there is no good reason to
assume the additional physics is universal, this may reintroduce the
flavor problem. Solutions to this problem have been proposed \cite{jack2}, for example
by combining this mechanism with an anomalous $U(1)$ gauge group
\cite{Jack:2000cd}.
\end{enumerate}

AMSB is a very characteristic corner of the model space of the
supersymmetry breaking. It could be important if the
tree level gravity mediation (which is generically one-loop order
larger than the AMSB piece) is suppressed. If AMSB is indeed the dominant contributor to the SUSY breaking, the suppression of other potential contributions (which are not suppressed in a generic supergravity Lagrangian) will be an important clue to the structure of the fundamental theory.

\subsection{Other benchmarks}

Although mSUGRA and AMSB are two of the most widely studied benchmark
scenarios, there are other benchmarks proposed in recent years with
different motivations. Here we give a brief survey of a few of
them which we will consider.

\subsubsection{Focus Point}

Focus point supersymmetry \cite{focuspoint}, in its simplest
realization, is a special corner of the 
mSUGRA parameter space. In this corner, very large scalar masses are
possible without violating naturalness constraints. The resulting low
energy parameters 
have a very important feature. Namely, the soft masses squared of the Higgs
boson, $m^2_{H_u}$, very important in electroweak symmetry
breaking, have pseudo fixed point behavior at the electroweak
scale. They can start with a wide range of input values and run to a
similar negative value at the low scale. This is interesting because it indicates that, in the
focus point region, electroweak symmetry breaking does not require fine-tuning in the high
energy input values. 

A typical feature of the focus point region is large scalar soft
masses (usually $\sim$ TeV). The LSP is usually a higgsino or a mixed higgsino-gaugino. This
is
quite different from the other cases where the LSP has a dominant
gaugino content. The main reason for a larger higgsino content is
again the larger input value of the soft scalar mass. The tree level 
electroweak symmetry breaking condition gives 
\begin{equation}
\frac{1}{2}m_Z^2 \sim - m_{H_u}^2 -\mu^2.
\end{equation}
In the typical mSUGRA scenarios, $m_{H_u}^2$ is driven to some large
negative value due to the RGE running. This requires a large value of
$\mu$ to give the correct Z mass. However, in the focus point region, it
is possible that the large input value of the scalar soft mass makes
$m_{H_u}^2$ less negative. Hence, a smaller value of $\mu$ is possible, which leads to
a larger higgsino content in the LSP.

%\subsubsection{Snowmass Benchmarks}

%At the 2001 Snowmass meeting, a set of benchmark supersymmetry
%scenarios were proposed \cite{snowmass}.  These benchmarks include
%representative models within mSUGRA, as well as gauge and anomaly
%mediated models. These selected cases help to illustrate the various
%phenomenological features associated with different areas of the large
%MSSM parameter space. We have considered the dark matter phenomenology
%for many of these selected benchmarks. {\bf liantao: read this section
%  and modify as you see fit} 

\subsubsection{Michigan Benchmarks}

Recently, another set of benchmark models has been proposed
\cite{michigan}. The main motivation of these scenarios is to consider different patterns of gaugino
masses (compared to the standard relation in
mSUGRA, for example). By doing so, the gluino mass does not need to be quite so heavy (as
implied by mSUGRA or AMSB relations). It is shown in Ref.~\cite{michigan} that the
fine-tuning of the electroweak symmetry breaking in the MSSM is dominated
by the gluino mass. Therefore,  a different prediction for the gluino
mass will generically offer hope of reducing fine-tuning.

Another distinctive feature of this set of benchmark models is that
each of them has a motivation in string theory, rather than
making arbitrary simplifying assumptions about the pattern of the soft
parameters at the input scale. 

Most of these proposed benchmarks fall in the category of gravity
mediated supersymmetry breaking. The main feature
of these constructions is the suppression of the tree level dilaton
contribution which give rise to mSUGRA-like universal input
values. Therefore, they are essentially different combinations of
1-loop contributions such as anomaly mediation, threshold corrections,
and Green-Schwarz counter terms. There are also two cases in which
the spectrum resembles the typical gauge mediated scenario but without a light gravitino. 

For more phenomenologically minded readers, the most important
information is the spectrum of superpartners presented in tables 1 and 2 of
Ref.~\cite{michigan}. 

\subsubsection{Gaugino Mediation}

Gaugino mediated supersymmetry breaking, proposed in
Ref.~\cite{Kaplan:1999ac}, represents another class of SUSY
breaking mediation motivated by the brane-world scenario. It achieves
the suppression of unwanted supersymmetry breaking effects, such as
the flavor violating couplings, by the separation of the observable
and hidden sectors via the separation of their respective branes. In the simplest
realization, it begins with a two brane set up. Most of the matter
fields, in particular the chiral fermions and their superpartners,
localize on one of those branes while the supersymmetry breaking
sector localizes on the other. If this is the whole description, it is again
similar to the anomaly mediation scenario. The new idea in this
scenario is that the vector multiplet, in particular, the gauginos,
are now allowed to propagate in the bulk. 

The pattern of soft masses in this scenario is straight forward to describe: 
\begin{enumerate}
\item The gauginos will have ``direct'' couplings to the supersymmetry
breaking sector. Therefore, their soft masses are proportional to $F/M$, where
$F$ is supersymmetry breaking order parameter and $M$ is some
underlying fundamental scale which characterizes the coupling between
gauginos and the hidden sector (since the coupling is usually of the
form of a non-renormalizable term suppressed by $M$). With proper
choices of $F$ and $M$, the gaugino masses in this scenario can be chosen to
be similar to any of the other supersymmetry breaking mediation
scenarios. 
\item The soft masses in this scenario cannot come from the direct
couplings due to the separation of the two branes. On the other hand,
they can be generated from the 1-loop diagrams in which a gaugino is
emitted, travels through the bulk to the supersymmetry breaking brane,
gets the information of SUSY breaking and then returns to join the
sfermion propagator again. Generically, the soft masses are then
suppressed relative to the gaugino mass by a loop factor,
$m^2_{\tilde{f}} \sim M_{\lambda}^2/(16 \pi^2)$. 
\item At low energy,  soft masses  and trilinear couplings generated
through the RGE runnings are all proportional to the gaugino masses
and are universal (since gauge coupling is flavor diagonal). Therefore, in
this scenario, any flavor violating effect is again suppressed. 
\end{enumerate}

Notice that the soft masses for sfermions are always positive in this
model since they essentially start from zero at the high scale and get
positive corrections from RGE running. 

In this paper, we have considered only a small sample of possible
gaugino mediated models in which the features of mSUGRA are manifest
while respecting the conditions of point 2 ($m^2_{\tilde{f}} \sim
M_{\lambda}^2/(16 \pi^2)$).

\section{Calculation of observable quantities and parameter scan}

\vspace{0.2in}

Assuming R-parity is conserved, a large portion of the parameter space in the
scenarios mentioned above can provide a neutralino which is a
viable dark matter candidate. The present relic density for such a particle can be calculated by
solving the Boltzman equations. This
calculation is more complicated, however, if the mass of another
supersymmetric particle is only slightly greater than that of the
LSP. In this situation, these particles may efficiently co-annihilate\footnote{Co-annihilations are annihilations
between LSPs and other supersymmetric particles, such as heavier neutralinos, charginos, staus, etc.},
lowering the relic abundance of a thermal relic
\cite{coannihilations}. We have used the DARKSUSY package for our relic
density calculation which includes all coannihilations with
charginos and neutralinos as well as all resonances and thresholds in
the cross section calculation \cite{darksusy}. To calculate the
supersymmetric particle spectrum, we have used the packages SUSPECT
\cite{suspect} and FeynHiggs \cite{feynhiggs}. 

The thermal relic density of neutralinos can be compared to the WMAP
result of $\Omega_{\rm{CDM}} h^2 = 0.113^{+0.016}_{-0.018}$
\cite{wmap}. It is possible, however, that the present density of dark matter particles may have been produced by non-thermal mechanisms
\cite{amsb,nonthermal}. With this in mind, we treat the WMAP dark
matter density measurement as an upper limit for the thermal relic
abundance of any given WIMP. For models which predict a thermal relic
density below the WMAP measurement, we assume that non-thermal
mechanisms are sufficient to generate the measured dark matter
density. 

There can be strong constraints on supersymmetric models imposed by
measurements of the $b \rightarrow s \gamma$ branching fraction
\cite{bsgamma}. In anomaly or gaugino mediated scenarios, the
supersymmetric flavor problem is solved naturally and this constraint
is typically not violated. In the general MSSM or mSUGRA, however,
many of the models violate this bound. On the other hand, we notice
that one should treat these constraints with caution since a small
deviation from mSUGRA, such as a small off diagonal term in the squark
mass matrix, could change significantly the prediction for $b
\rightarrow s \gamma$ without effecting the dark matter phenomenology
significantly. 

Ref~\cite{suspect} also present a scenario, called the
``phenomenological'' MSSM, which covers a subset of the full MSSM
parameter space (18 free parameters). It essentially relaxes the universality conditions in
mSUGRA between the first two generations and the third, the two Higgs
soft masses squared, and the three gaugino masses. All of the off diagonal
entries in the sfermion mass matrices as well as the CP violating
phases are still set to be zero\footnote{Notice that CP violating
  phases could have important impact on the dark matter
  phenomenology. For example, they could significantly change the
  gaugino-quark-squark coupling and thereby have a significant impact both on
  annihilation and elastic scattering cross sections \cite{Brhlik:2000dm}.}. Although
this ``truncated'' 
parameterization of the MSSM is not specifically linked to any
supersymmetry breaking mechanism and is not general enough to give a model independent conclusion, it is useful to illustrate what the dark
matter signature may look like if the real supersymmetry breaking
scenario is outside of the models we have considered. Therefore, we
also include this more general scenario in our study.

At the 2001 Snowmass meeting, a set of benchmark supersymmetry
scenarios were proposed \cite{snowmass}.  These benchmarks include
representative models within mSUGRA, as well as gauge and anomaly
mediated models. These selected cases help to illustrate the various
phenomenological features associated with different areas of the large
MSSM parameter space. Although in principle they are already included
in the parameter scan of mSUGRA and AMSB, it is interesting to
consider them separately as representative corners of the parameter
space. Therefore, we have studied the dark matter phenomenology
for many of these selected benchmarks and included them in our figures.

\section{Results and Discussion}

\vspace{0.2in}

\subsection{Results For Direct Detection Experiments}

\vspace{0.2in}

The rates and sensitivities for direct detection experiments can be
reliably calculated with relatively small uncertainties which result from
measurements of the local dark matter density. For this reason,
results from direct detection experiments are among the most useful
tests of supersymmetric dark matter. 

\begin{figure}
\centering\leavevmode
\epsfxsize=6in\epsffile{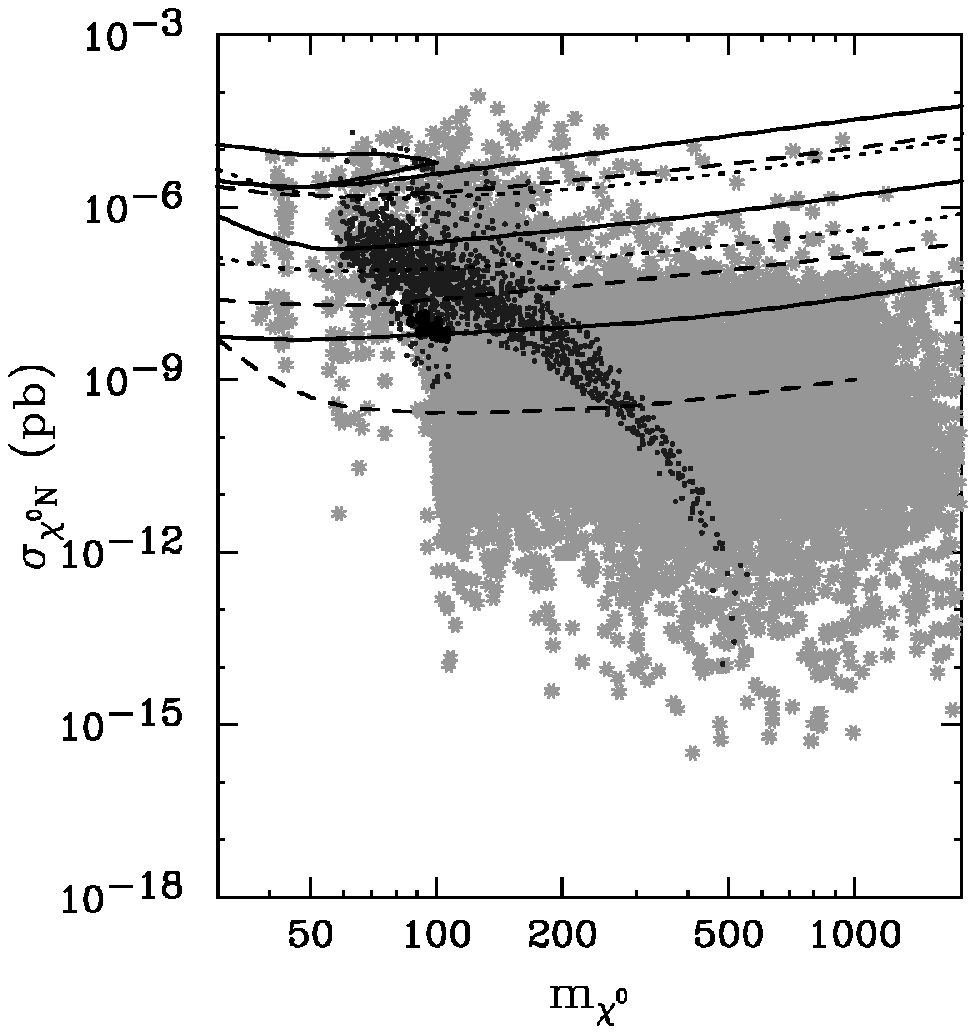}

\caption[]{
The scalar (spin-independent) neutralino-nucleon elastic scattering cross section verses neutralino mass. Each point corresponds to a theoretical prediction for a specific SUSY model. The light shaded area corresponds to a general or phenomenological MSSM model. The darker colored region are those models limited to the constrained MSSM (mSUGRA). Also shown as dark circles are the points which we describe as gaugino mediated
supersymmetry breaking models. See the text for more details. For each point shown, the relic density is below
the maximum value allowed by the WMAP data ($\Omega_{\chi} h^2 \le 0.129$). Models that violate accelerator limits are not shown, except for $b$ to $s\gamma$ limits which will be discussed later. The experimental limits and sensitivities, shown as lines, from top to bottom (on the right) are CDMS-March 2002 (solid) \cite{cdmsnow}, ZEPLIN 1-final 2002 (dashed) \cite{zeplin1}, Edelweiss-2000+2002 (dots) \cite{edelweiss}, CRESST II-projected limit (solid) \cite{crest}, CDMS-projected limit (dashed) \cite{cdms}, Edelweiss II-projected limit (dots)
\cite{edelweiss2}, ZEPLIN 4-projection (solid) \cite{zeplin4} and XENON-1 ton, projected limit (dashed) \cite{xenon}. Also shown, as a solid contour, is the region which the DAMA collaboration claims discovery \cite{dama}. For a summary of present limits and projections for all direct dark matter experiments, see Ref.~\cite{alldirect}.
\label{fig:one}
}\end{figure}

\begin{figure}
\centering\leavevmode
\epsfxsize=6in\epsffile{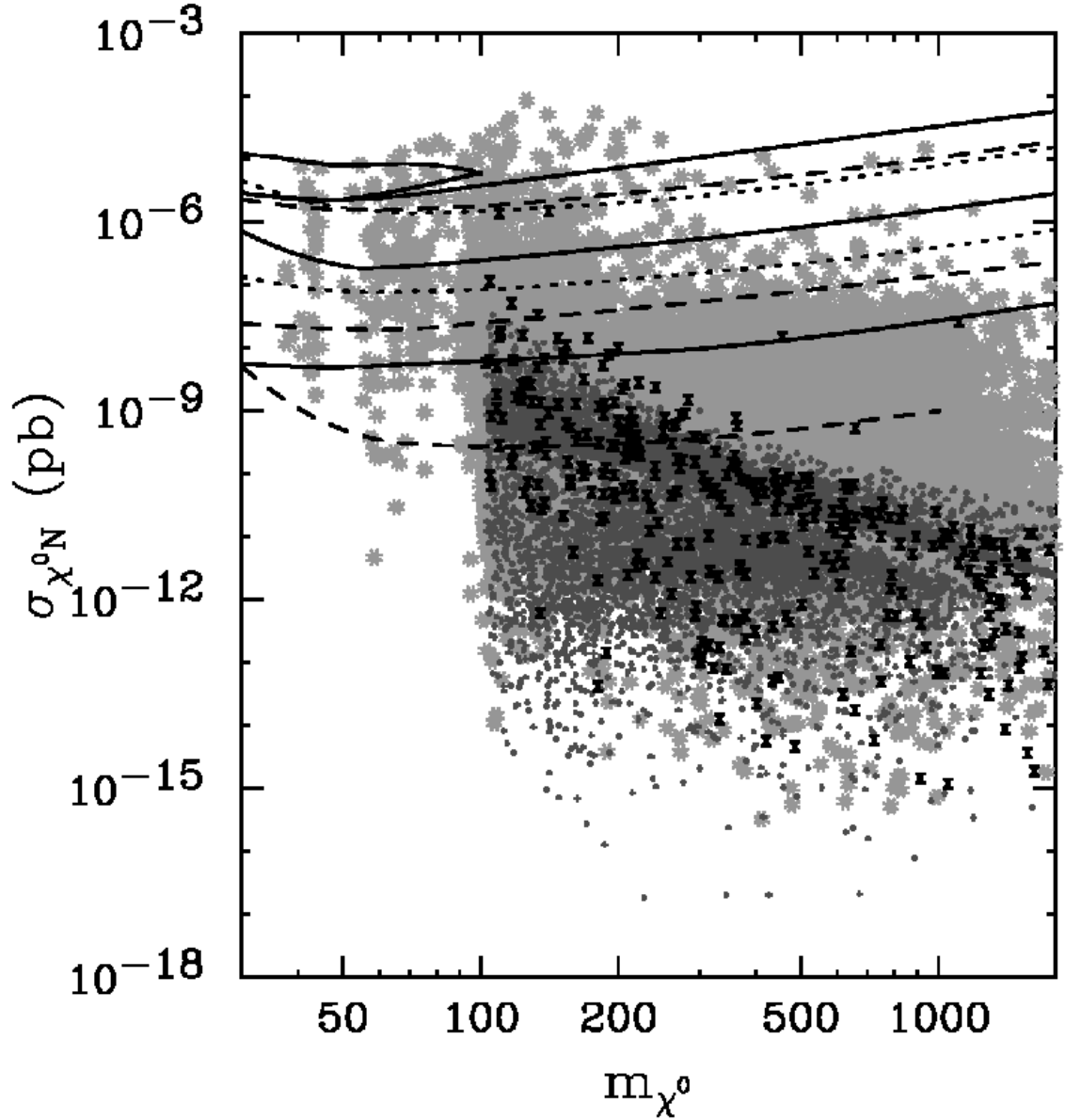}

\caption[]{
The scalar (spin-independent) neutralino-nucleon elastic scattering cross section verses neutralino mass. Each point
corresponds to a theoretical prediction for a specific SUSY model. The light shaded area corresponds to a general or phenomenological MSSM model. The darker shaded region shows those models limited to minimal Anomaly Mediated Supersymmetry Breaking (mAMSB) models. Also shown are black X's which represent non-minimal AMSB models. See the text for more details. For each point shown, the relic density is below the maximum value allowed by the WMAP data ($\Omega_{\chi} h^2 \le 0.129$). Models that violate accelerator limits are also not shown, except for $b$ to $s\gamma$ limits which will be discussed later. The experimental limits and sensitivities, shown as lines, from top to bottom (on the right) are CDMS-March 2002 (solid) \cite{cdmsnow}, ZEPLIN 1-final 2002 (dashed) \cite{zeplin1}, Edelweiss-2000+2002 (dots) \cite{edelweiss}, CRESST II-projected limit (solid) \cite{crest}, CDMS-projected limit (dashed) \cite{cdms}, Edelweiss II-projected limit (dots) \cite{edelweiss2}, ZEPLIN 4-projection (solid) \cite{zeplin4} and XENON-1 ton, projected limit (dashed) \cite{xenon}. Also shown, as a solid contour, is the region which the DAMA collaboration claims discovery \cite{dama}. For a summary of present limits and projections for all direct dark matter experiments, see Ref.~\cite{alldirect}.
\label{fig:two}
}\end{figure}

In figures~\ref{fig:one} and~\ref{fig:two}, we show comparisons of the
neutralino mass to the scalar (spin independent) neutralino-nucleon
elastic scattering cross section for a variety of models. In figure 1,
the lightly shaded area corresponds to a the phenomenological
MSSM search described in the previous section. The darker shaded region represents those models limited to
minimal supergravity (mSUGRA). For this case, we have allowed $m_0$,
$m_{1/2}$ and $A_0$ to vary between 0 and 10,000 GeV. We selected
$\tan \beta$ to have values in the range of 1 to 50 and have allowed
$\mu$ to take on either positive or negative values. We acknowledge that 10,000 GeV is probably outside of the interesting range for $m_{1/2}$ or $A_0$, but we choose to allow such values in an attempt to be inclusive. Our results would vary only slightly if we were to use an upper limit of a few TeV, for example. Also shown as
dark circles are the points which we consider to be gaugino mediated
supersymmetry breaking models. In such models, $m_0$ is less than
$4\pi\,m_{1/2}$. In this manifestation, gaugino mediation is a subset
of mSUGRA.  

For each point shown, the relic density is below the maximum value
allowed by the WMAP data ($\Omega_{\chi} h^2 \le 0.129$). Models that
violated accelerator limits are also not shown, except for $b$ to
$s\gamma$ limits which may be violated. The experimental limits and
sensitivities, shown as lines, from top to bottom (on the right) are
CDMS-March 2002 (solid) \cite{cdmsnow}, ZEPLIN 1-final 2002 (dashed)
\cite{zeplin1}, Edelweiss-2000+2002 (dots) \cite{edelweiss}, CRESST
II-projected limit (solid) \cite{crest}, CDMS-projected limit (dots)
\cite{cdms}, Edelweiss II-projected limit (dashed) \cite{edelweiss2},
ZEPLIN 4-projection (solid) \cite{zeplin4} and XENON-1 ton projected
limit (dashed) \cite{xenon}. Also shown, as a solid contour, is the
region which the DAMA collaboration claims evidence \cite{dama}. For a
summary of present limits and projections for all direct dark matter
experiments, see Ref.~\cite{alldirect}. 

Figure~\ref{fig:one} indicates that very little of mSUGRA space has
been effectively probed by existing direct searches.  However, future
experiments such as GENIUS and XENON will have the ability to probe
nearly all mSUGRA models with LSP masses below about 200 GeV. 

In figure~\ref{fig:two}, again the lightly shaded area corresponds to the phenomenological MSSM.  The darker shaded
region shows those models limited to minimal Anomaly Mediated Supersymmetry
Breaking (mAMSB) models. In such models, we allowed $m_0$ to vary
between 0 and 10,000 GeV and the gravitino mass, $m_{3/2}$, to vary
between 25,000 and 3,000,000 GeV. Again $\tan \beta$ was selected
between 1 and 50 and $\mu$ is allowed to have either sign. We acknowledge that 10,000 GeV or 3,000,000 GeV are probably outside of the interesting range for $m_{0}$ or $m_{3/2}$, respectively, but we choose to allow such values in an attempt to be inclusive. Our results would vary only slightly if we were to use a smaller upper limit.

We assumed
universality of the scalar masses in the minimal AMSB model.  Also
shown in figure~\ref{fig:two} are black X's which represent randomly
selected non-minimal AMSB models. In these models, in addition to the
parameters varied in the minimal AMSB, the scalar masses squared were
allowed to vary from $m^2_0$ at the GUT scale. Each scalar mass squared was set to $m^2_0$ multiplied by a factor
randomly
selected between 5 and -5.

Figure~\ref{fig:two} shows that even with future experiments, AMSB
scenarios are very difficult to test with direct detection methods. In
non-minimal models, these prospects are slightly better. A primary
reason direct detection is difficult in AMSM models is that the value
of $M_3$ is considerably higher than $M_2$ or $M_1$, and, therefore,
squark masses tend to be much larger than the LSP mass. This leads to
suppression in the s-channel diagram for neutralino-quark
scattering. Additionally, AMSB typically predicts a gaugino-like LSP,
less favorable for direct detection than a mixed gaugino-higgsino or pure higgsino. 

\begin{figure}                                                                                                                   
\centering\leavevmode                                                                                                            
\epsfxsize=6in\epsffile{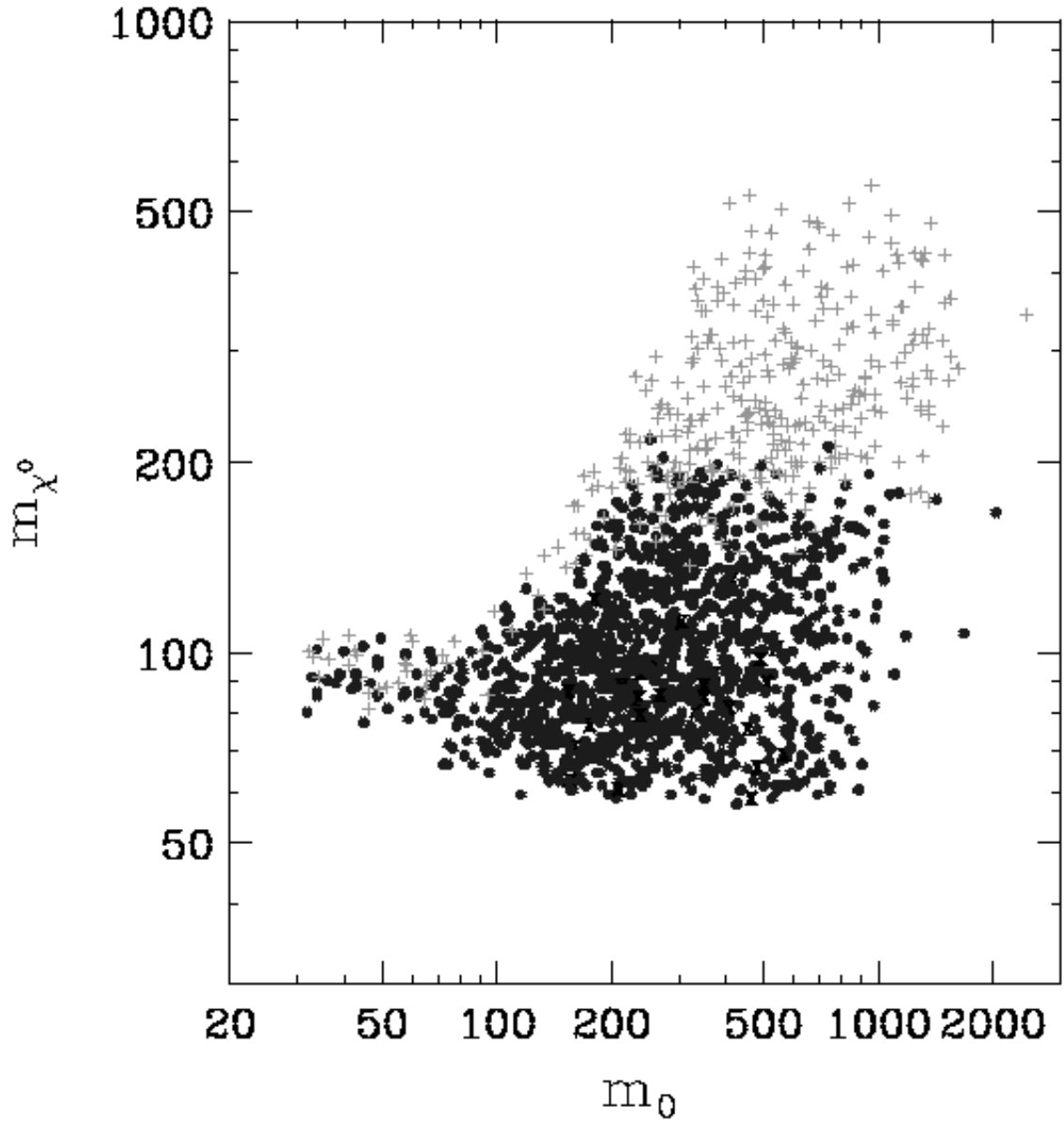}                                      
                                                                                                                                 
\caption[]{                                                                                                                      
The mass of the lightest neutralino versus the universal scalar mass, $m_0$, in
the mSUGRA scenario.  The darkest points are those currently excluded by one or more of CDMS-March 2002, ZEPLIN 1-final 2002 or
Edelweiss-2000+2002. The intermediate points are testable by planned experiments including CRESST II, CDMS (Soudan),
Edelweiss II or ZEPLIN 4. The lightest points are beyond the sensitivity of ZEPLIN 4. The           
models shown meet the requirements of the previous figures.                                                          
\label{fig:mm0direct}                                                                                                                  
}\end{figure}

Figure~\ref{fig:mm0direct}, illustrating the effect of scalars in direct detection, shows the mass of the lightest
neutralino versus the universal scalar mass, $m_0$, in
the mSUGRA scenario.  The darkest points are those currently excluded by either CDMS-March 2002, ZEPLIN 1-final 2002 or          
Edelweiss-2000+2002. The intermediate points are testable by planned experiments including CRESST II, CDMS (Soudan),             
Edelweiss II or ZEPLIN 4. The lightest points are beyond the sensitivity of ZEPLIN 4. 

\begin{figure}
\centering\leavevmode
\epsfxsize=6in\epsffile{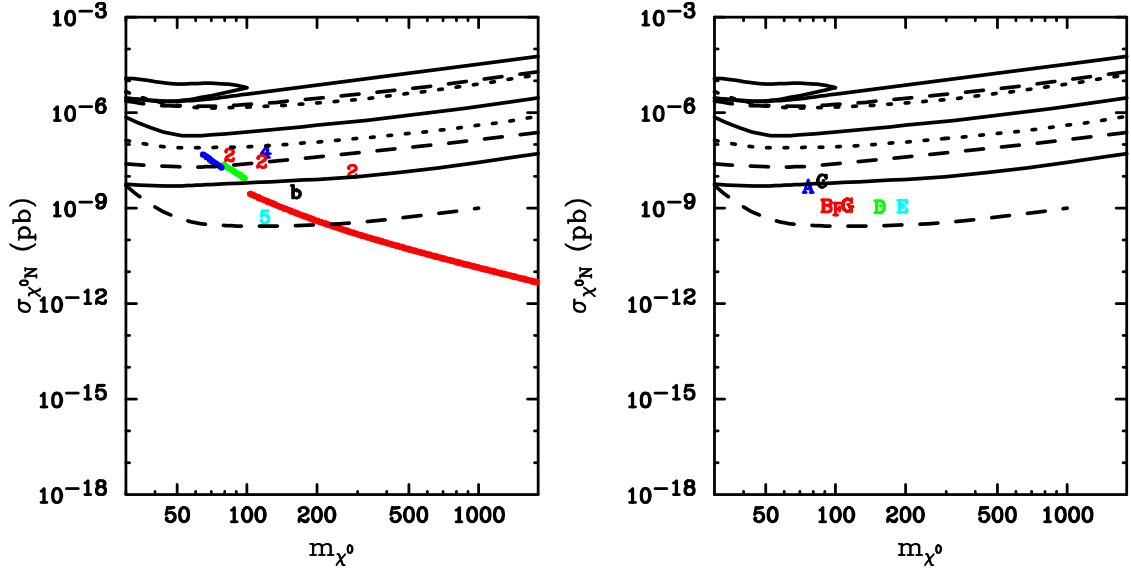}

\caption[]{
The scalar (spin-independent) neutralino-nucleon elastic scattering cross section verses neutralino mass. Each point corresponds to a theoretical prediction for a specific Snowmass (left frame) or Michigan (right frame) benchmark model.

In the left frame, Snowmass slopes 1a, 3 and 9 are shown as lines from left to right, respectively. Snowmass points 1b, and samples of 2, 4 and 5 are shown as B, 2, 4 and 5 in the figure. In the right frame, Michigan benchmark models A through G are shown. See the text for more details.

Along the length of each Snowmass slope shown, the relic density is below the maximum value allowed by the WMAP data ($\Omega_{\chi} h^2 \le 0.129$). However, some of the points (rather than slopes) shown produce a larger relic density. The experimental limits and sensitivities, shown as lines, from top to bottom (on the right) are CDMS-March 2002 (solid) \cite{cdmsnow}, ZEPLIN 1-final 2002 (dashed) \cite{zeplin1}, Edelweiss-2000+2002 (dots) \cite{edelweiss}, CRESST II-projected limit (solid) \cite{crest}, CDMS-projected limit (dashed) \cite{cdms}, Edelweiss II-projected limit (dots) \cite{edelweiss2}, ZEPLIN 4-projection (solid) \cite{zeplin4} and XENON-1 ton, projected limit (dashed) \cite{xenon}. Also shown, as a solid contour, is the region which the DAMA collaboration claims discovery \cite{dama}. For a summary of present limits and projections for all direct dark matter experiments, see Ref.~\cite{alldirect}.
\label{fig:twob}
}\end{figure}

Figure~\ref{fig:twob} shows the sensitivities to
direct experiments for a variety of Snowmass and Michigan benchmark
scenarios. Snowmass slopes 1a, 3 and 9 and shown as lines from left to
right, respectively. Snowmass point 1b, and samples of slopes 2, 4 and 5, are
shown as b, 2, 4 and 5 in the figure. For the Michigan benchmark
scenarios, models A through G are shown. 

Along the length of each slope shown (in the Snowmass plot), the relic
density is below the maximum value allowed by the WMAP data
($\Omega_{\chi} h^2 \le 0.129$). However, some of the points (rather
than slopes) shown produce a larger relic density.

\subsection{Results For Gamma-Ray Experiments}

\vspace{0.2in}

In the case of a cuspy or spiked galactic halo, gamma-ray astronomy
can provide a strong test of supersymmetric dark matter. For models
with flat galactic cores, however, such observations are nearly
impossible in planned experiments. 

\begin{figure}
\centering\leavevmode
\epsfxsize=6in\epsffile{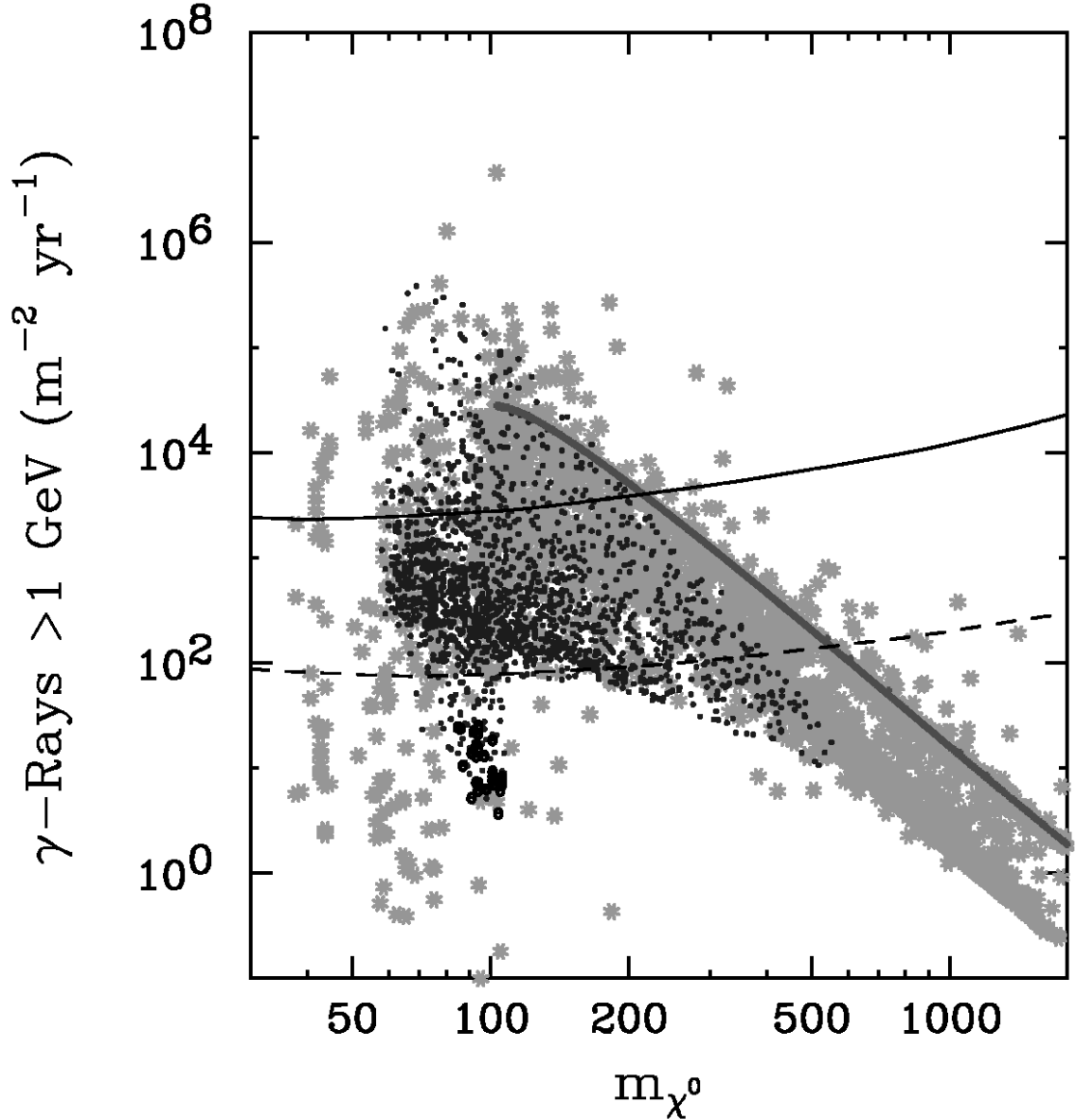}

\caption[]{
The rate of continuum gamma-rays above 1 GeV per square meter, per year of exposure, from the galactic center, verses neutralino mass. A smooth NFW halo profile is assumed. If a Moore, {\it et. al.} model were considered, each point would produce approximately a factor of $10^3$ more events. As in figures~\ref{fig:one} and~\ref{fig:two}, the lightly shaded region is for the general or phenomenological MSSM, the darker region corresponds to mSUGRA models, the shaded line represents AMSB models (minimal and non-minimal) and black circles indicate gaugino mediated models. For each point shown, the relic density is below the maximum value allowed by the WMAP data ($\Omega_{\chi} h^2 \le 0.129$). Models that violate accelerator limits are also not shown, except for $b$ to $s\gamma$ limits which will be discussed later. See the text for more details.

Also shown are limits from EGRET (solid) and the predicted sensitivity for the future experiment GLAST (dashed) \cite{brenda}. Note that EGRET is already sensitive to some models, and GLAST will be sentivity to many models, especially for those models with a light neutralino. In AMSB scenarios, the mass parameters $M_1$ and $M_2$ are proportional to their respective couplings, so the neutralino mass and annihilation cross section become correlated, leading to the single line across the figure. 
\label{fig:three}
}\end{figure}

In figure~\ref{fig:three}, the rate of continuum gamma-rays above 1
GeV per square meter, per year of exposure, from the galactic center
is shown verses neutralino mass. A smooth NFW halo profile is
assumed. If a Moore, {\it et. al.} model were considered, each point
would produce approximately a factor of $10^3$ more events; considerably
more if a spiked halo model were used. As in figures~\ref{fig:one}
and~\ref{fig:two}, the lightly shaded region represents the phenomenological MSSM, the darker region corresponds to mSUGRA models,
and the darker line represents the AMSB models (minimal and
non-minimal).  Also shown are limits from EGRET (solid) and the predicted
sensitivity for the future experiment GLAST (dashed)
\cite{brenda}. Note that EGRET is already sensitive to some models,
and GLAST will be sensitive to many models, especially for those
models with a relatively light neutralino. In AMSB scenarios, the mass parameters
$M_1$ and $M_2$ are proportional to their respective couplings, so the
neutralino mass and annihilation cross section become correlated,
leading to the single line across the figure.  

\begin{figure}                                                                                                                   
\centering\leavevmode                                                                                                            
\epsfxsize=6in\epsffile{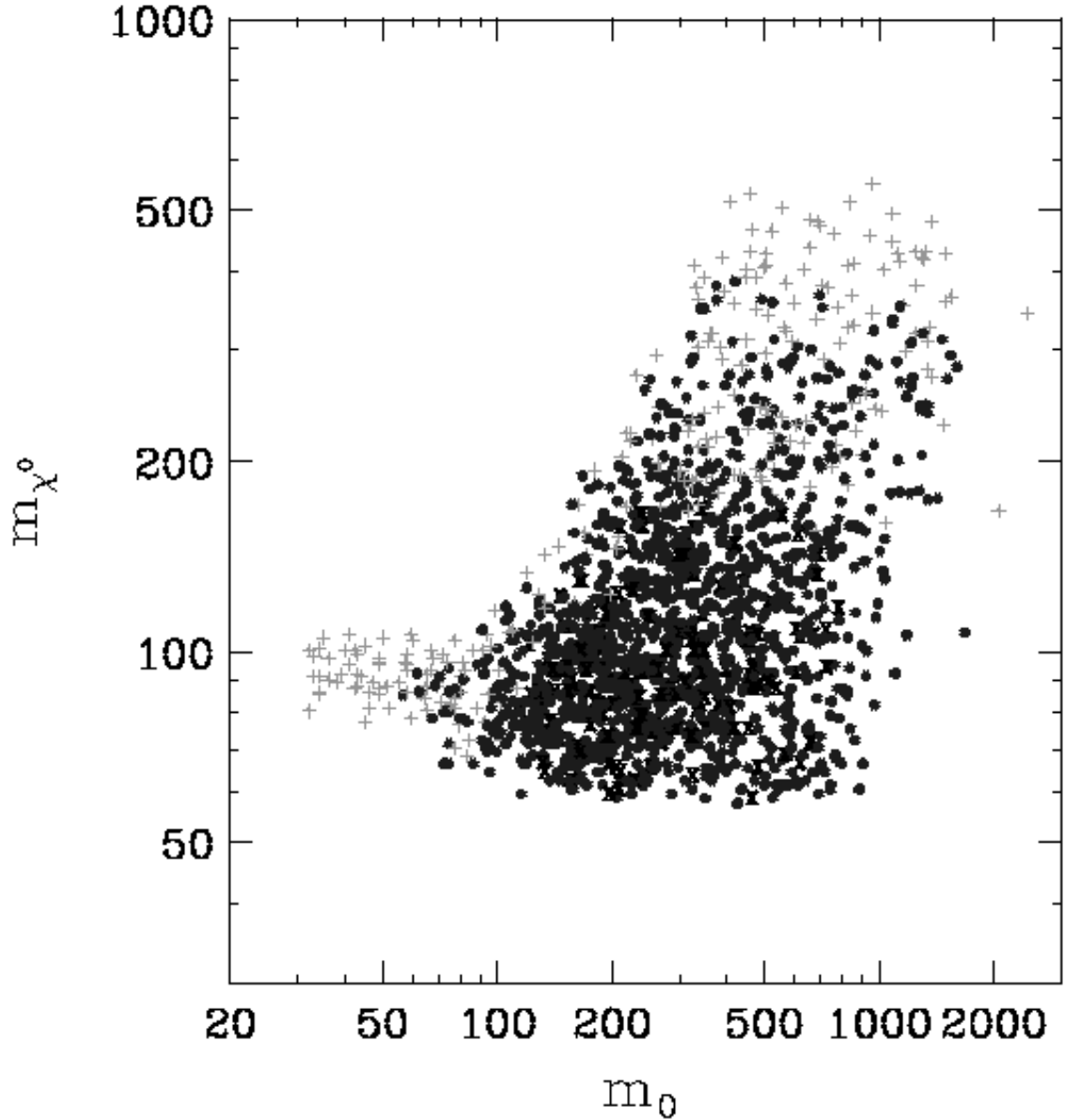}
\caption[]{                                                                                                                      
The mass of the lightest neutralino versus the universal scalar mass, $m_0$, in        
the mSUGRA scenario.  The darkest points are those currently excluded by EGRET assuming a smooth NFW halo profile. The
intermediate points are testable by the GLAST experiment. The lightest points are beyond the sensitivity of GLAST. The
models shown meet the requirements of the previous figure.                                            
\label{fig:mm0gamma}
}\end{figure}

Figure~\ref{fig:mm0gamma}, illustrating the effect of scalars in gamma-ray production, shows the mass of the lightest
neutralino versus the universal scalar mass, $m_0$, in                                                  
the mSUGRA scenario.  The darkest points are those currently excluded by EGRET assuming a smooth NFW halo profile. The           
intermediate points are testable by the GLAST experiment. The lightest points are beyond the sensitivity of GLAST. 

\begin{figure}
\centering\leavevmode
\epsfxsize=6in\epsffile{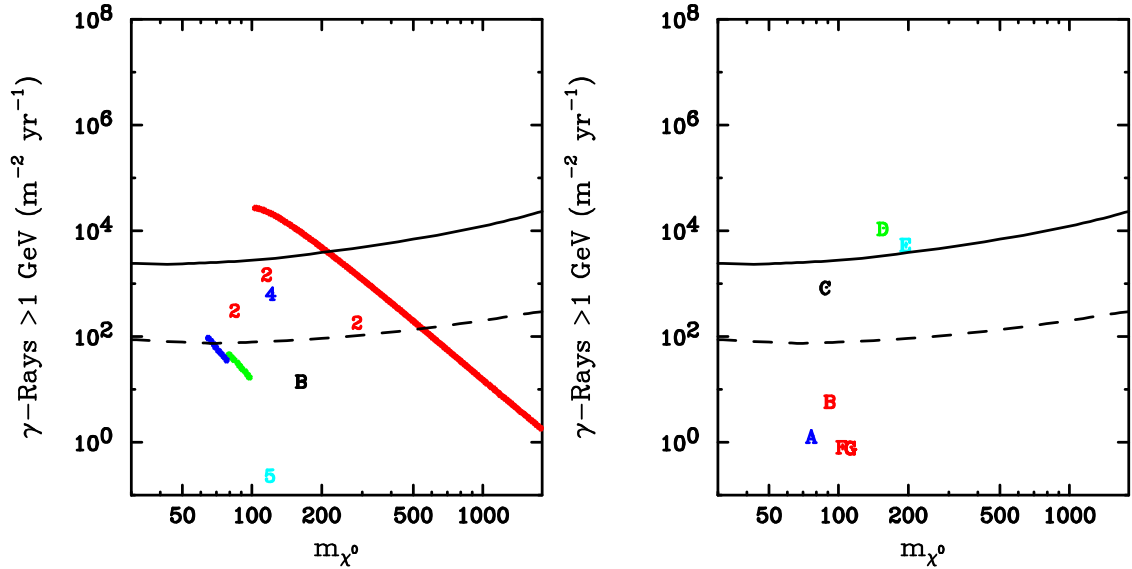}

\caption[]{
The rate of continuum gamma-rays above 1 GeV per square meter, per year of exposure, from the galactic center, verses neutralino mass. A smooth NFW halo profile is assumed. If a Moore, {\it et. al.} model were considered, each point would produce approximately a factor of $10^3$ more events.

In the left frame, Snowmass slopes 1a, 3 and 9 are shown as lines from left to right, respectively. Snowmass points 1b, and samples of 2, 4 and 5 are shown as B, 2, 4 and 5 in the figure.  In the right frame, each point corresponds to a theoretical prediction for a specific Michigan benchmark model. Models A through G are shown. See the text for more details.

Along the length of each slope shown, the relic density is below the maximum value allowed by the WMAP data ($\Omega_{\chi} h^2 \le 0.129$). However, some of the points (rather than slopes) shown produce a larger relic density. 

Also shown are limits from EGRET (solid) and predicted sensitivity for the future experiment GLAST (dashed) \cite{brenda}.
\label{fig:threeb}
}\end{figure}

Again, we also show results for Snowmass and Michigan benchmark
scenarios. Figure~\ref{fig:threeb} shows the
sensitivities to gamma-ray experiments for these models. The models shown are the same as those shown for the direct
experiments. 

\begin{figure}
\centering\leavevmode
\epsfxsize=6in\epsffile{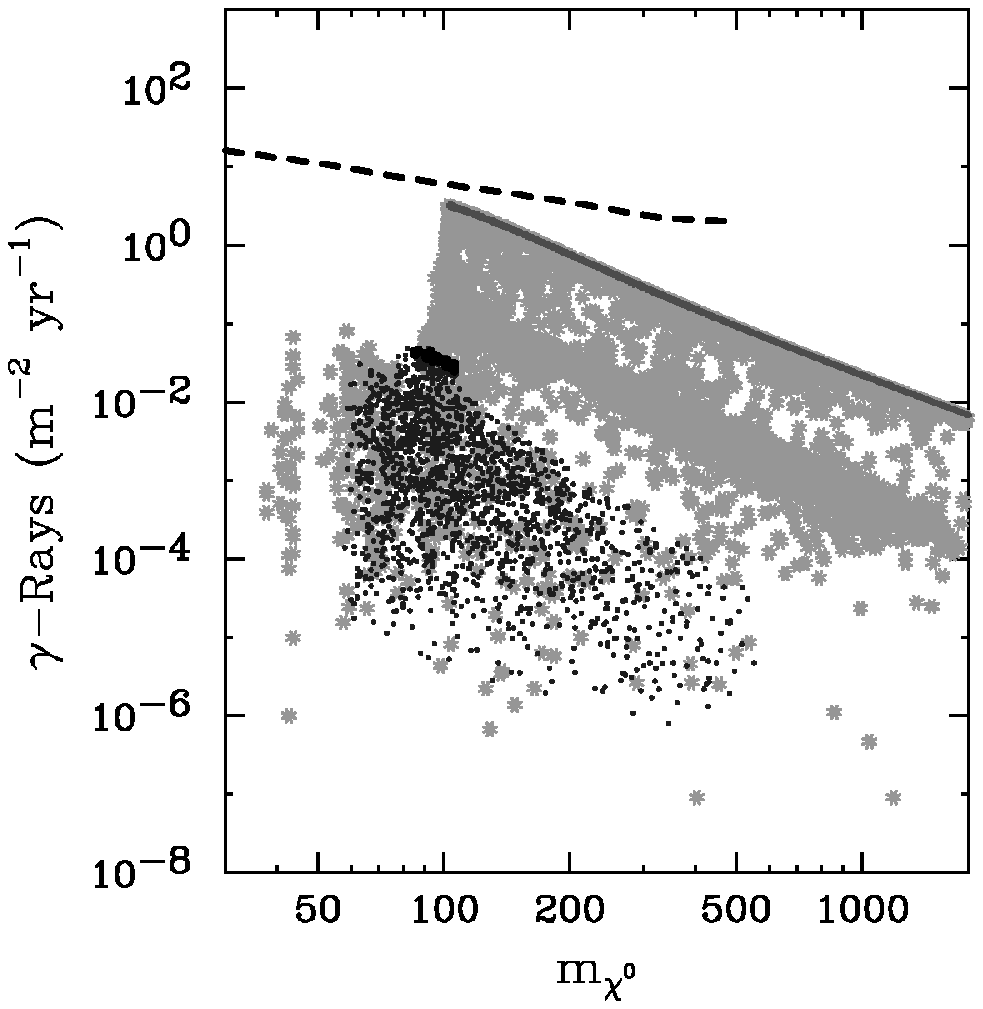}

\caption[]{
The rate of gamma-rays from the process $\chi^0 \chi^0 \rightarrow \gamma \gamma$ per square meter, per year of exposure, from the galactic center, verses neutralino mass. A smooth NFW halo profile is assumed. If a Moore, {\it et. al.} model were considered, each point would produce approximately a factor of $10^3$ more events. As in figures~\ref{fig:one} and~\ref{fig:two}, the lightly shaded region is for the general or phenomenological MSSM, the darker region corresponds to mSUGRA models, the shaded line represents AMSB models (minimal and non-minimal) and black circles indicate gaugino mediated models. For each point shown, the relic density is below the maximum value allowed by the WMAP data ($\Omega_{\chi} h^2 \le 0.129$). Models that violate accelerator limits are also not shown, except for $b$ to $s\gamma$ limits which will be discussed later. See the text for more details.

Also shown is the predicted sensitivity for the future experiment GLAST (dashed) \cite{bergstrom}. In AMSB scenarios, the mass parameters $M_1$ and $M_2$ are proportional to their respective couplings, so the neutralino mass and annihilation cross section become correlated, leading to the single line across the figure. 
\label{fig:four}
}\end{figure}

\begin{figure}
\centering\leavevmode
\epsfxsize=6in\epsffile{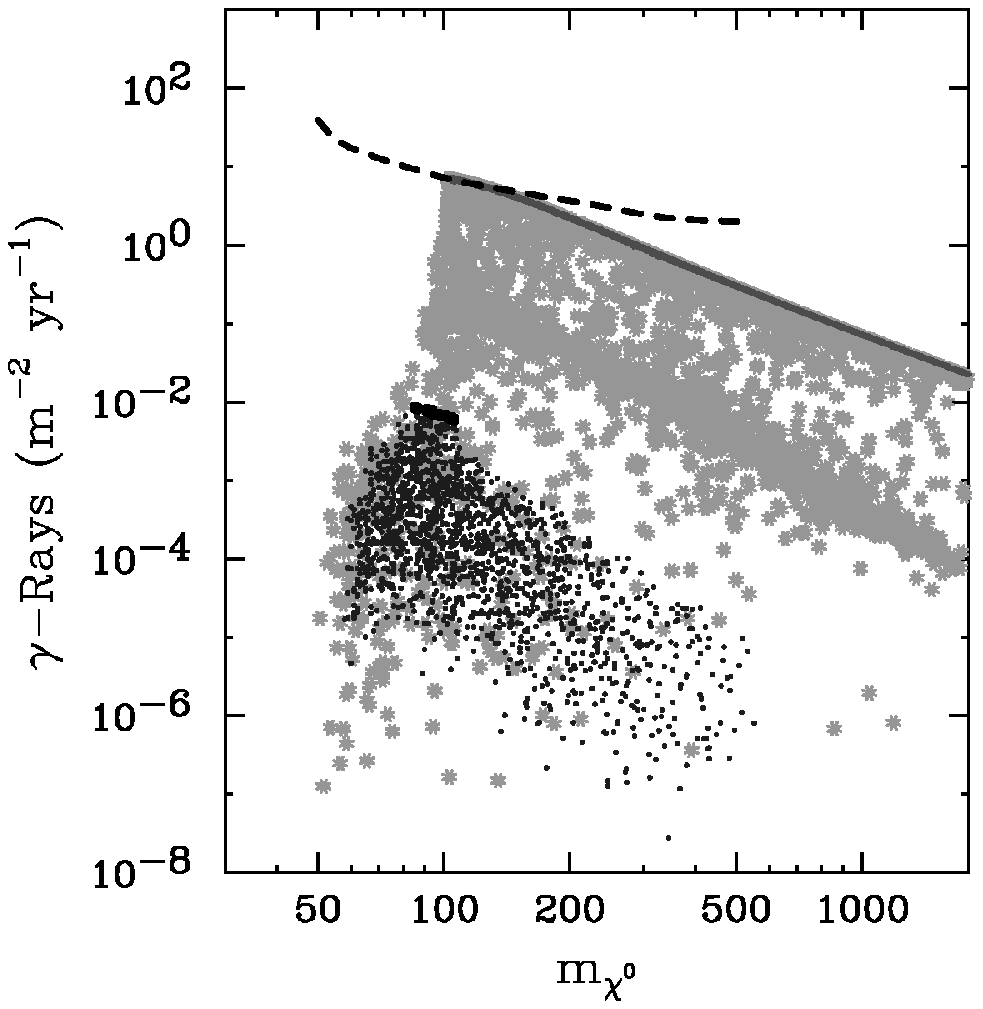}

\caption[]{
The rate of gamma-rays from the process $\chi^0 \chi^0 \rightarrow \gamma Z$ per square meter, per year of exposure, from the galactic center, verses neutralino mass. A smooth NFW halo profile is assumed. If a Moore, {\it et. al.} model were considered, each point would produce approximately a factor of $10^3$ more events. As in figures~\ref{fig:one} and~\ref{fig:two}, the lightly shaded region is for the general or phenomenological MSSM, the darker region corresponds to mSUGRA models, the shaded line represents AMSB models (minimal and non-minimal) and black circles indicate gaugino mediated models. For each point shown, the relic density is below the maximum value allowed by the WMAP data ($\Omega_{\chi} h^2 \le 0.129$). Models that violate accelerator limits are also not shown, except for $b$ to $s\gamma$ limits which will be discussed later. See the text for more details.

Also shown is predicted sensitivity for the future experiment GLAST (dashed) \cite{bergstrom}. In AMSB scenarios, the mass parameters $M_1$ and $M_2$ are proportional to their respective couplings, so the neutralino mass and annihilation cross section become correlated, leading to the single line across the figure.
\label{fig:five}
}\end{figure}

In figures ~\ref{fig:four} and ~\ref{fig:five}, the rate of gamma-rays from the line producing processes, $\chi^0 \chi^0 \rightarrow \gamma \gamma$ and $\chi \chi
\rightarrow \gamma Z$, are shown \cite{bergstrom}. The
sensitivity of GLAST is shown (dashed) for comparison. Again a smooth
NFW halo profile is assumed. AMSB scenarios are the best prospect for
line detection, although if a more cuspy model, such as Moore, {\it
  et. al.}, were realized, even some mSUGRA models may be observable.

\subsection{Results For Neutrino Experiments}

\vspace{0.2in}

Neutrinos from WIMP annihilation in the Sun (or Earth), like direct
detection, provide a method of dark matter detection with rates and
sensitivities which can be reliably calculated. 

\begin{figure}
\centering\leavevmode
\epsfxsize=6in\epsffile{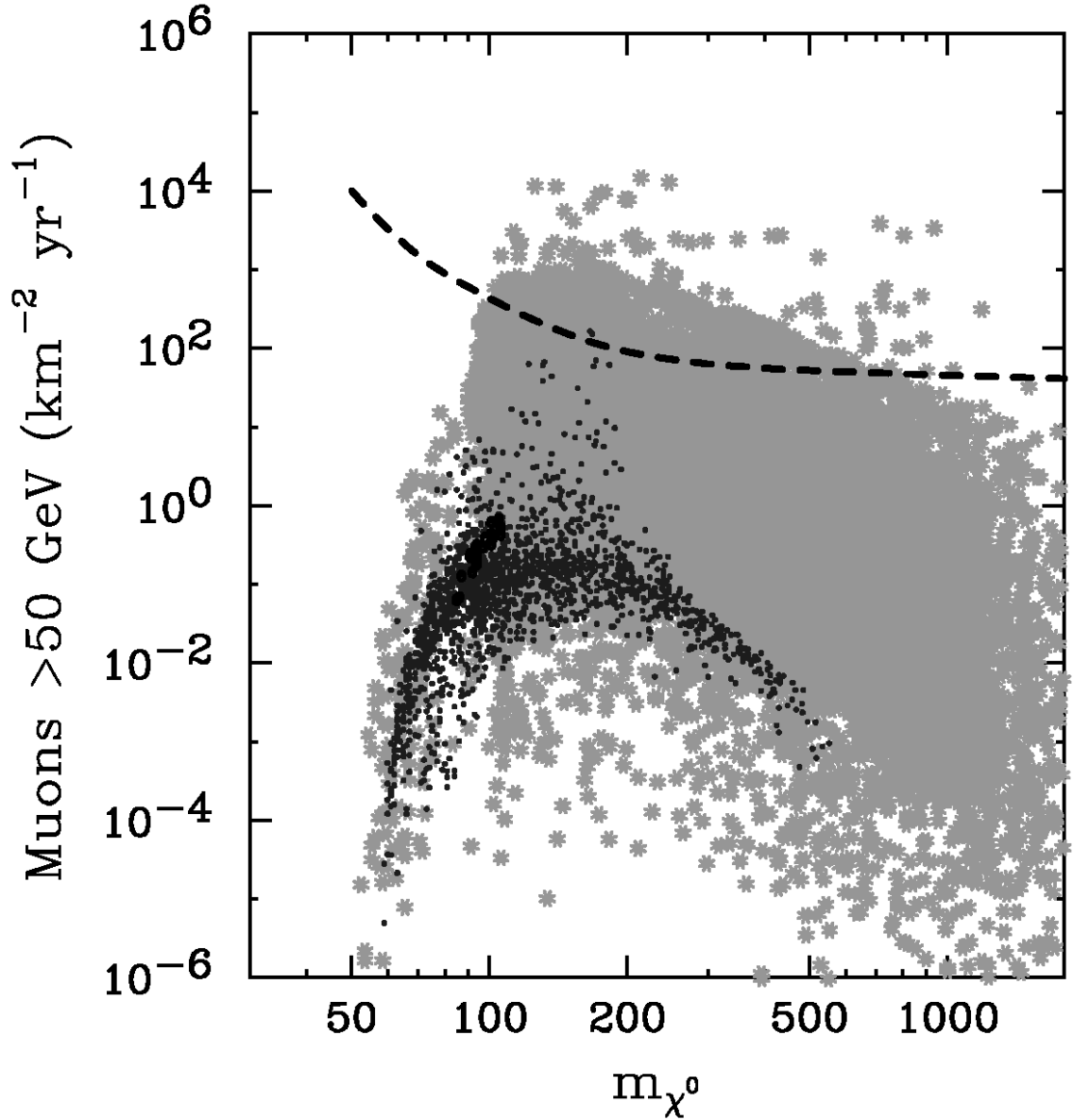}

\caption[]{
The rate of muons (from charged current muon neutrino interactions) above 50 GeV per square kilometer, per year, from neutralino annihilation in the sun, verses neutralino mass. As in figure~\ref{fig:one}, the lightly shaded region represents the general or phenomenological MSSM, the darker region corresponds to mSUGRA models and black circles indicate gaugino mediated models. For each point shown, the relic density is below the maximum value allowed by the WMAP data ($\Omega_{\chi} h^2 \le 0.129$). Models that violate accelerator limits are also not shown, except for $b$ to $s\gamma$ limits which will be discussed later. See the text for more details.

Also shown is the predicted sensitivity for the next generation neutrino telescope IceCube (dashed) \cite{edsjolimits}. 
\label{fig:six}
}\end{figure}

\begin{figure}
\centering\leavevmode
\epsfxsize=6in\epsffile{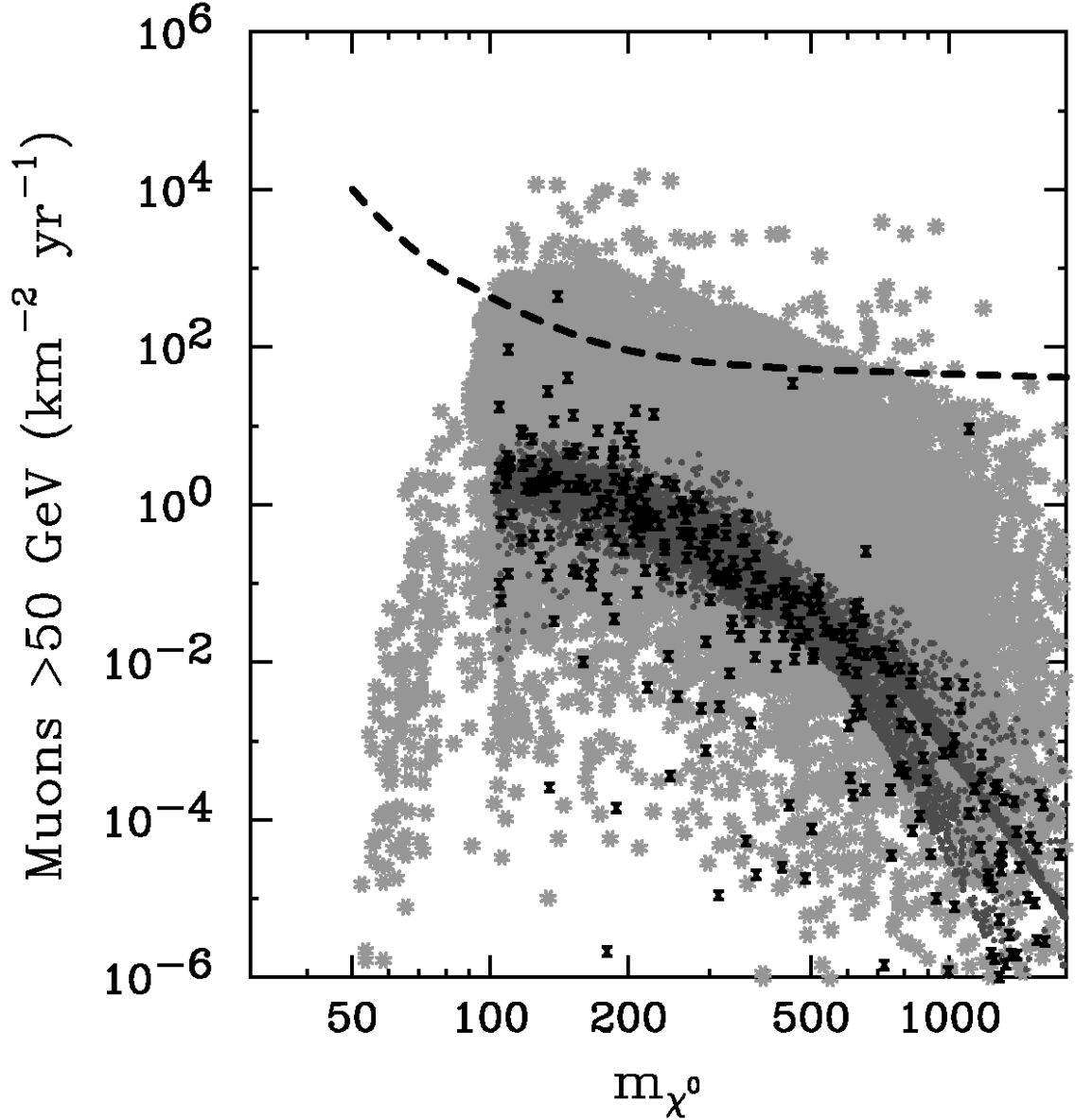}

\caption[]{
The rate of muons (from charged current muon neutrino interactions) above 50 GeV per square kilometer, per year, from neutralino annihilation in the sun, verses neutralino mass. As in figure~\ref{fig:two}, the lightly shaded region is for the general or phenomenological MSSM, the darker shaded region show those models limited to minimal Anomaly Mediated Supersymmetry Breaking (mAMSB) models. Also shown are black X's which represent non-minimal AMSB models. For each point shown, the relic density is below the maximum value allowed by the WMAP data ($\Omega_{\chi} h^2 \le 0.129$). Models that violate accelerator limits are also not shown, except for $b$ to $s\gamma$ limits which will be discussed later. See the text for more details. Also shown is the predicted sensitivity for the next generation neutrino telescope IceCube (dashed)
\cite{edsjolimits}. 
\label{fig:seven}
}\end{figure}

In figures~\ref{fig:six} and~\ref{fig:seven}, we show the rates in a
kilometer scale neutrino telescope, such as IceCube, per year from
neutralino annihilation in the Sun. The models are represented by the
various types of shading as in figures~\ref{fig:one}
and~\ref{fig:two}. The sensitivity projected for the IceCube
experiment (dashed) is also shown \cite{edsjolimits}.  A 50 GeV muon
energy threshold has been imposed.  

Some models in the phenomenological MSSM
with a substantial higgsino fraction can have very large scattering
cross sections and, therefore, provide observable neutrino
rates. On the other hand, in either mSUGRA or AMSB scenarios, few
models can be probed with 
kilometer scale neutrino telescopes. An important exception is the
foucs point region of the mSUGRA parameter space, in which the LSP
also has a large higgsino component. These rates do not depend
strongly on the choice of halo 
profile as is the case for observations of the galactic center.

\begin{figure}
\centering\leavevmode
\epsfxsize=6in\epsffile{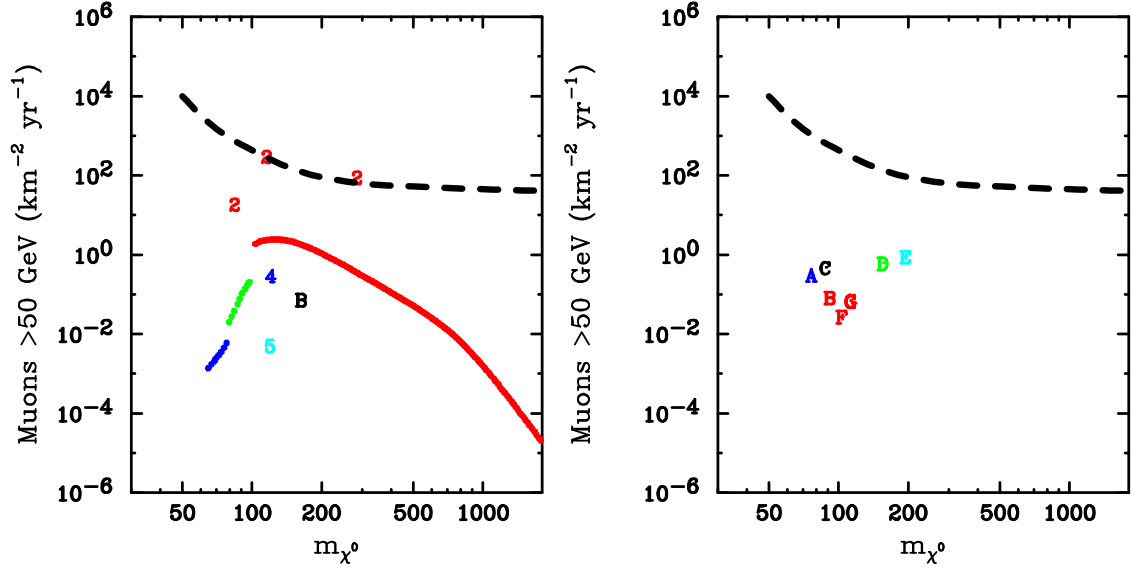}

\caption[]{
The rate of muons (from charged current muon neutrino interactions) above 50 GeV per square kilometer, per year, from
neutralino annihilation in the sun, verses neutralino mass. 

In the left frame, Snowmass slopes 1a, 3 and 9 and shown as lines from left to right, respectively. Snowmass points 1b, and samples of 2, 4 and 5
are shown as B, 2, 4 and 5 in the figure. In the right frame, each point corresponds to a theoretical prediction for a specific Michigan benchmark model. Models A through G are shown. See the text for more details. 

Along the length of each Snowmass slope shown, the relic density is below the maximum value allowed by the WMAP data ($\Omega_{\chi}
h^2 \le 0.129$). However, some of the points (rather than slopes) shown produce a larger relic density. Also shown is the predicted sensitivity for the next generation neutrino telescope IceCube (dashed) \cite{edsjolimits}.
\label{fig:sevenb}
}\end{figure}

Again, we also show results for Snowmass and Michigan benchmark
scenarios. Figure~\ref{fig:sevenb} shows the
sensitivities of neutrino experiments for these models. The models
shown are the same as for the direct and gamma-ray experiments. 

\subsection{Results For Positron and Anti-Proton Experiments}

\vspace{0.2in}

\begin{figure}
\centering\leavevmode
\epsfxsize=6in\epsffile{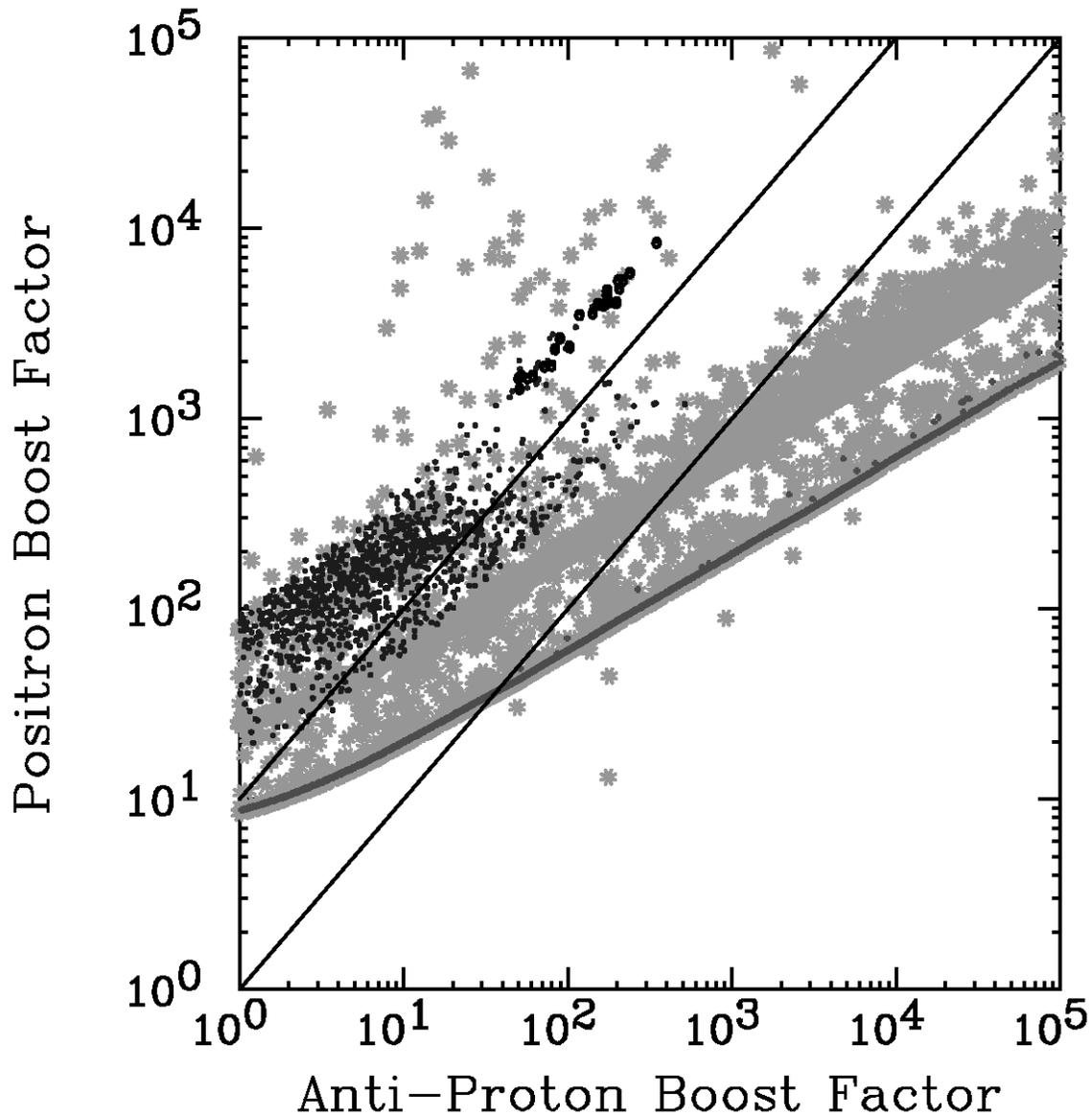}

\caption[]{
The positron boost factor required to explain the HEAT positron excess \cite{heat1995,heat2000} versus the maximum
anti-proton boost factor consistent with the BESS anti-proton cosmic ray measurement \cite{bess}. As in
figures~\ref{fig:one} and~\ref{fig:two}, the lightly shaded region is for the general or phenomenological MSSM, the darker region corresponds to mSUGRA models, the shaded line corresponds to AMSB models (minimal and non-minimal) and black circles indicate gaugino mediated models. For each point shown, the relic density is below the maximum value allowed by the WMAP data ($\Omega_{\chi} h^2 \le 0.129$). Models that violate accelerator limits are also not shown, except for $b$ to $s\gamma$ limits which will be discussed later. See the text for more details. Solid lines correspond to the case when the positron boost factor is equal to the maximum anti-proton boost factor and for the case when the positron boost factor ten times larger than the maximum 

anti-proton boost factor. See the text for further discussion.
\label{fig:eight}
}\end{figure}

Figure~\ref{fig:eight} compares the positron boost factor required to
explain the HEAT data to the maximum anti-proton boost factor which
does not violate the BESS bound. Results are shown for a variety of
models and SUSY breaking scenarios. The models are represented by the
various types of shading as in figure~\ref{fig:three}. It is somewhat
difficult to interpret these results, however, due to large
uncertainties in the galactic halo model. First, the question of how
large the positron boost factor could be is a subject of some
debate. It is known that on the very large scales of galaxies and
clusters of galaxies, such enhancement factors can exceed $\sim 10^2$
according to simulations \cite{largescale}. Perhaps boost factors significantly above this scale could be considered unrealistic. 

A second uncertainty is the degree to which the positron and
anti-proton boost factors could differ. A ratio of 1 to
10 between the two factors is certainly reasonable, but it is
difficult to determine how much larger this ratio could be.  In
figure~\ref{fig:eight}, lines are shown representing the cases of
equal positron and anti-proton boost factors and a positron boost
factor ten times the anti-proton boost factor. Perhaps models to the
upper-left of this second line are less likely to be capable of
explaining the observed positron excess. 

Note that even for a smooth halo profile (boost factors of one), a
fraction of the supersymmetry models considered produced an
anti-proton flux larger than would be consistent with the BESS data. 

An LSP with lesser Bino content, such as in the AMSB (wino) or focus point (higgsino) scenarios, will have a greater chance of producing the observed positron excess. Such an LSP may annihilate dominantly into W bosons (if kinematically allowed) which makes the positron spectrum considerably easier to fit. Also, not that our positron analysis is not a fit to the observed complete positron spectrum, but rather only a fit to a single point (near the observed peak), disregarding the spectral shape. We also assume the background positron flux is known \cite{positronbg}. A more careful analysis should be done using a $\chi^2$ fit to the entire observed spectrum, with the normalization of the background positron flux treated as a free parameter. The result of such an analysis \cite{positronswang} will be that there are supersymmetric models capable of producing the excess without a prohibitively large boost factor. Our analysis presented here is qualitatively right in the sense that it compares the relative potential for producing a positron excess between different models, which is what we are focusing on here.

\subsection{Correlations Between Different Experimental Searches and
  Constraints}

In some cases, the neutralino characteristics which determine the
observability of one detection method also play an important role for
another method. It is useful to identify these features in order to
discuss the prospects for neutralino dark matter detection over
multiple experimental methods. 

\begin{figure}
\centering\leavevmode
\epsfxsize=6in\epsffile{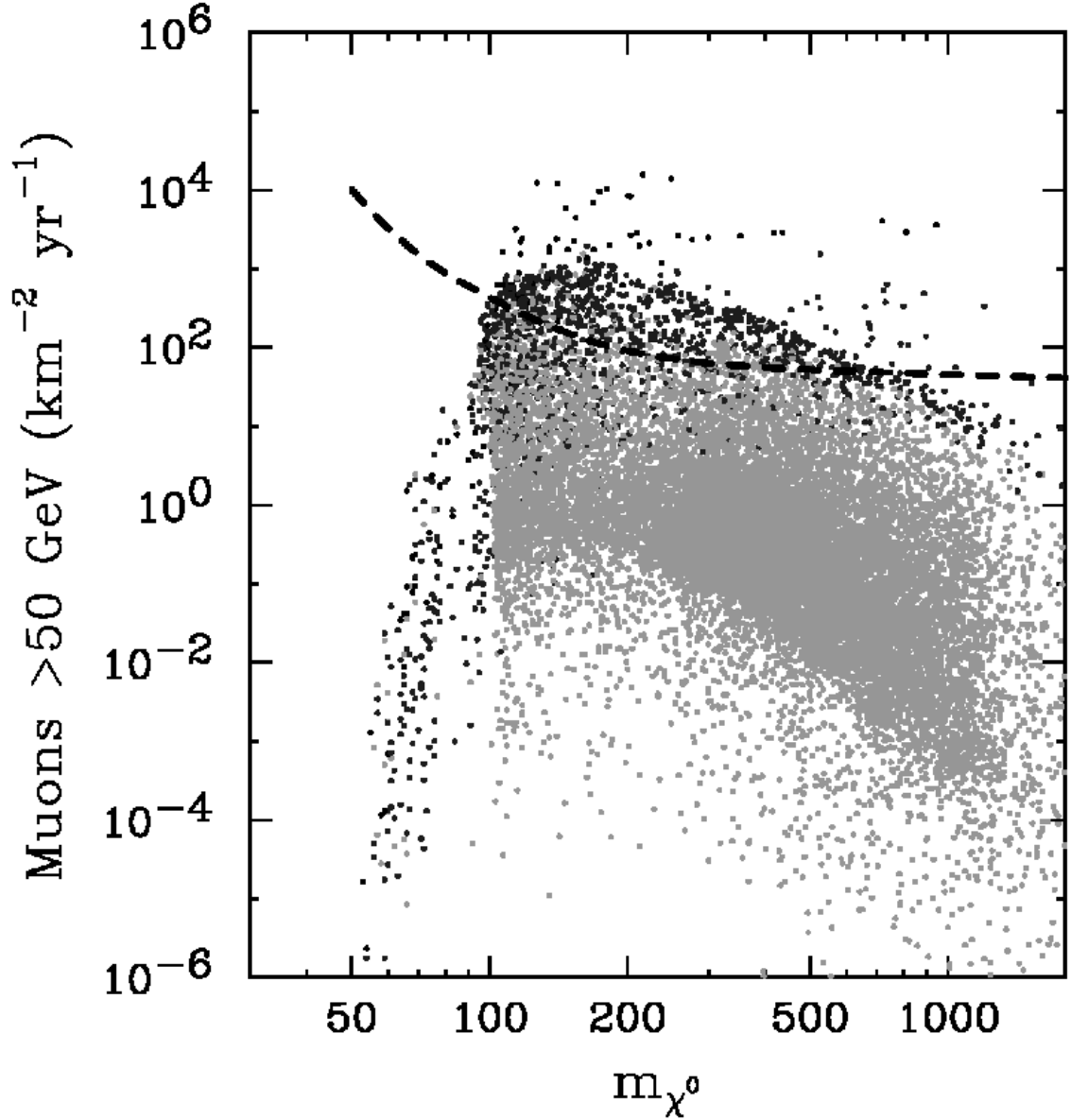}

\caption[]{
The rate of muons (from charged current muon neutrino interactions) above 50 GeV per square kilometer, per year, from neutralino annihilation in the sun, verses neutralino mass. All models show were randomly selected from the general or phenomenological MSSM, as described in the text. Black X's represent models which have allready been excluded by current experiments such as CDMS-March 2002 \cite{cdmsnow}, ZEPLIN 1-final 2002 \cite{zeplin1} and Edelweiss-2000+2002 \cite{edelweiss}. Darker points represent

 those models which can be tested by planned experiments with a sensitivity near that of the ZEPLIN 4-projection
\cite{zeplin4}. Lighter points fall below this sensitivity. Also shown is the predicted sensitivity for the next
generation neutrino telescope IceCube (dashed) \cite{edsjolimits}. For each point shown, the relic density is below the
maximum value allowed by the WMAP data ($\Omega_{\chi} h^2 \le 0.129$). Models that violate accelerator limits are also not shown, except for $b$ to $s\gamma$ limits which will be discussed later. See the text for more details.
\label{fig:eleven}
}\end{figure}

Prospects for both direct detection experiments and indirect detection experiments which rely on WIMP scattering on the Sun or Earth depend
on the elastic scattering cross section of WIMPs with nucleons. For
this reason, the flux of neutrinos from the Sun from neutralino
annihilation is related to the event rate in direct detection
experiments. Figure~\ref{fig:eleven} shows the rate of muons (from
charged current neutrino interactions) from neutralino annihilation in
the Sun for the phenomenological MSSM. Black points represent
models which have allready been excluded by current experiments such
as CDMS-March 2002 \cite{cdmsnow}, ZEPLIN 1-final 2002 \cite{zeplin1}
and Edelweiss-2000+2002 \cite{edelweiss}. Darker points represent
those models which can be tested by planned experiments with a
sensitivity near that of the ZEPLIN 4-projection
\cite{zeplin4}. Lighter points fall below this sensitivity.  

There is a clear correlation between the sensitivities of these two
types of experiments. For models with a neutralino mass substantially
above the energy threshold of neutrino telescopes, the
sensitivity of an experiment such as IceCube is very similar to that
of next generation direct detection experiments. Given this
observation, these two methods of detection will likely be capable of
confirming an observation made by the other technique. Note that this
relationship is not present for all SUSY breaking scenarios,
however. 

\begin{figure}
\centering\leavevmode
\epsfxsize=6in\epsffile{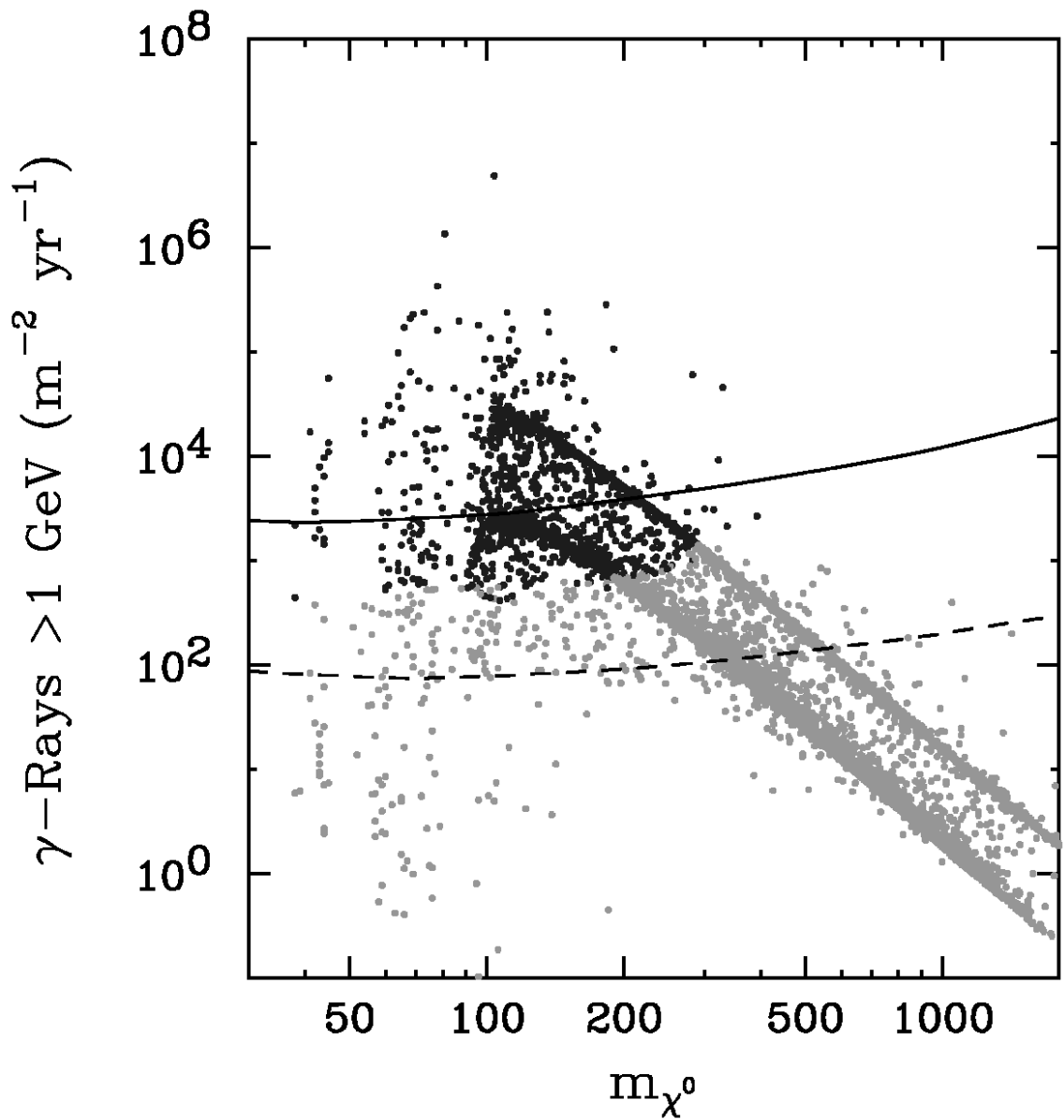}

\caption[]{
The rate of continuum gamma-rays above 1 GeV per square meter, per year of exposure, from the galactic center, verses neutralino mass. A smooth NFW halo profile is assumed. If a Moore, {\it et. al.} model were considered, each point would produce approximately a factor of $10^3$ more events. Models in the lightly shaded region do not overproduce anti-protons in the case of a
smooth halo profile, as measured with the BESS experiment \cite{bess}. Models in the darker region, however, require some
degree of fine tuning in the halo model, i.e. placing clumps at large distances, to accomodate this limit. Each model shown
corresponds to a point in the general or phenomenological MSSM, as described in the text. For each point shown, the relic density is below the maximum value allowed by the WMAP data ($\Omega_{\chi} h^2 \le 0.129$). Models that violate accelerator limits are also not shown, except for $b$ to $s\gamma$ limits. See the text for more details. Also shown are
limits from EGRET (solid) and predicted sensitivity for the future experiment GLAST (dashed) \cite{brenda}. Note that those models in which EGRET is sensitive, anti-proton measurements appear to disfavor.
\label{fig:thirteen}
}\end{figure}

Several methods of indirect detection depend strongly on the
neutralino-neutralino annihilation cross section. For this reason,
event rates for these experiments are correlated. For example, in
figure~\ref{fig:thirteen}, we show the rate of gamma-rays above 1 GeV,
per square meter, per year of exposure, from the galactic center,
verses neutralino mass, as in figure~\ref{fig:three}. Particularly
high neutralino annihilation cross sections will not only result in a
high rate of gamma-rays, but also a high rate of other annihilation
products, including anti-protons from the galactic halo. Models in the
lightly shaded region do not overproduce anti-protons in the case of a
smooth halo profile, as measured with the BESS experiment
\cite{bess}. Models in the darker region, however, require some degree
of fine tuning in the halo model, i.e. placing clumps at large
distances, to accomodate this limit. The consequence of this
comparison is the conclusion that for those models in which EGRET is
sensitive, they are also disfavored by anti-proton measurements. If,
however, a more cuspy halo model is considered, the rate from
annihilations in the galactic center region will increase, although
the anti-proton and positron rates, which depend primarily on the local halo characteristics, will not be
significantly affected.  

\begin{figure}
\centering\leavevmode
\epsfxsize=6in\epsffile{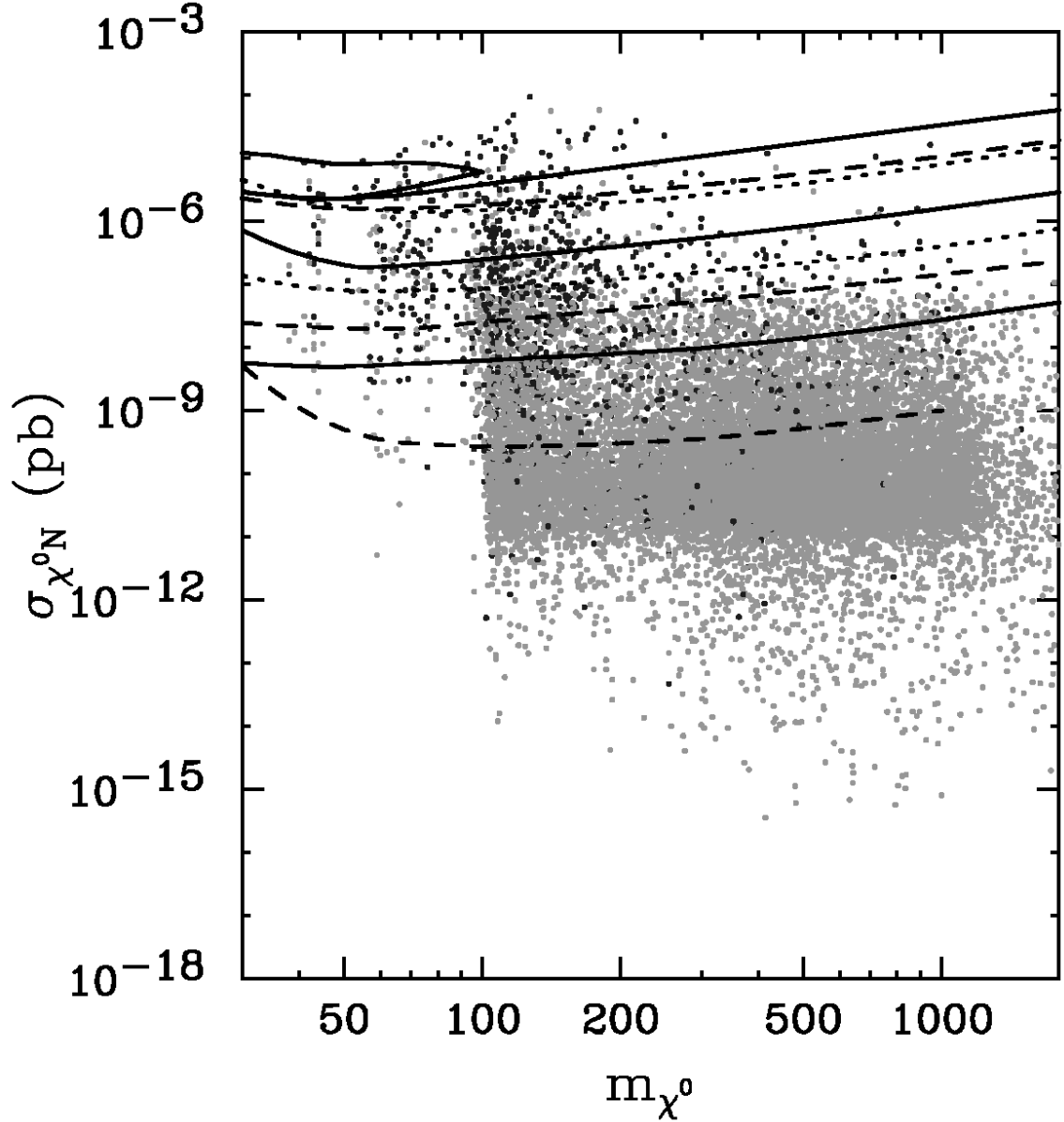}

\caption[]{
The impact of the $b \rightarrow s \gamma$ constraint \cite{bsgamma} on direct detection for models in the general or phenomenological MSSM. Lighter points are models which do not violate this constraint. Darker points do violate this constraint. Also shown are various experimental limits and sensitivities as in figures~\ref{fig:one} and~\ref{fig:two}. See the text for more details.
\label{fig:nine}
}\end{figure}

\begin{figure}
\centering\leavevmode
\epsfxsize=6in\epsffile{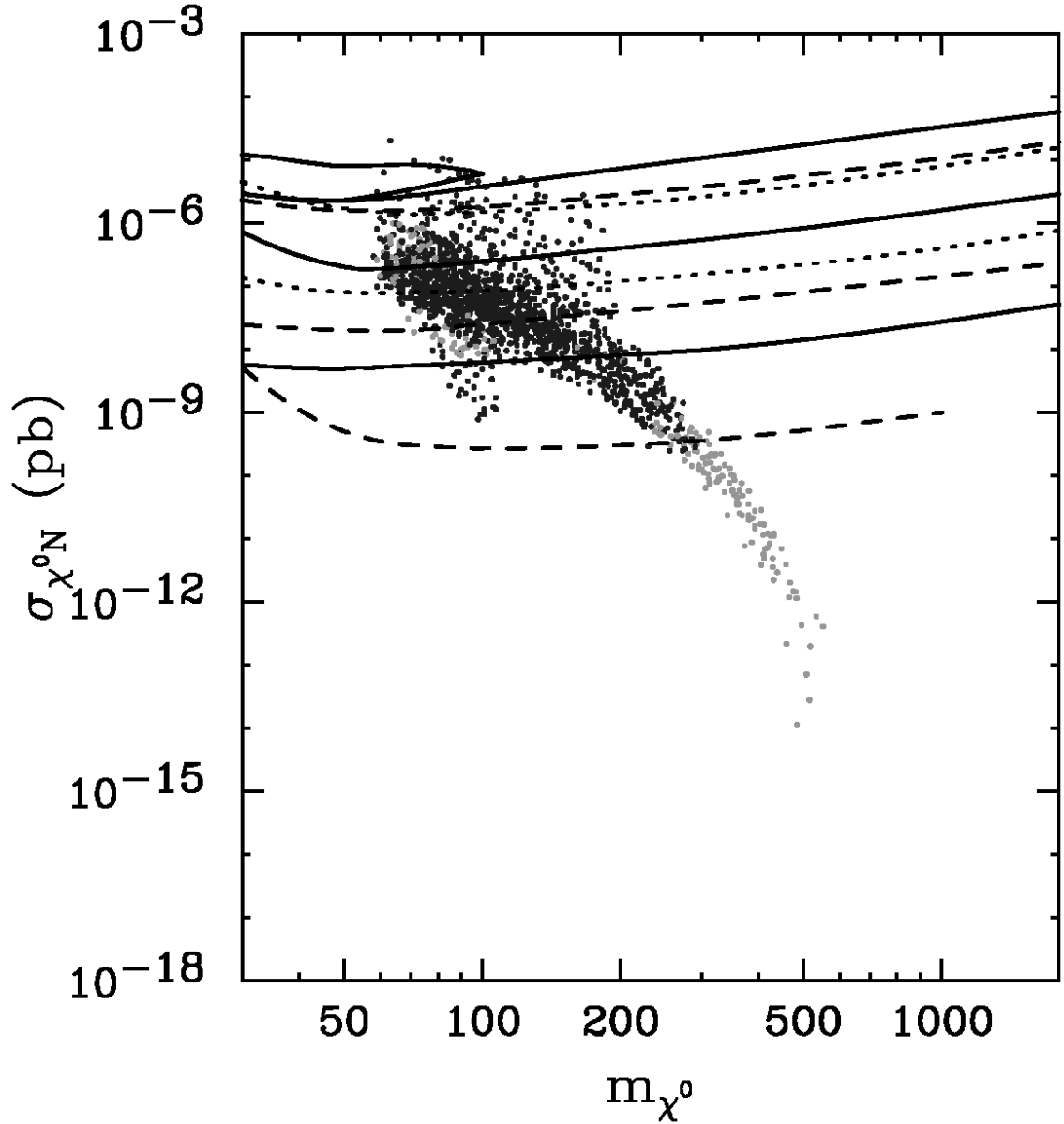}

\caption[]{
The impact of the $b \rightarrow s \gamma$ constraint \cite{bsgamma} on direct detection for models in the mSUGRA scenario. Lighter points are models which do not violate this constraint. Darker points do violate this constraint. Note that the small cluster of lighter points near 60-100 GeV and $10^{-8}$ pb are those mSUGRA points which we consider gaugino mediated ($m^2_{\tilde{f}} \sim
M_{\lambda}^2/(16 \pi^2)$). Also shown are various experimental limits and sensitivities as in figures~\ref{fig:one} and~\ref{fig:two}. See the text for more details.
\label{fig:ten}
}\end{figure}

If constraints from the measurements of the $b \rightarrow s \gamma$
branching fraction are considered, many models in the mSUGRA scenario
can be strongly disfavored. In figures~\ref{fig:nine}
and~\ref{fig:ten}, we show the impact of the $b \rightarrow s \gamma$
constraint on the prospects for direct detection. In
figure~\ref{fig:nine}, all sample points shown correspond to the phenomenological MSSM, as described
earlier. Figure~\ref{fig:ten} shows sample points in the mSUGRA
scenario. In each figure, lighter points do not violate the $b
\rightarrow s \gamma$ constraint. Darker points do violate this
constraint. 

Note that those models most easily tested by direct detection
experiments are more likely to violate the $b \rightarrow s \gamma$
constraint. This is especially apparent in the case of mSUGRA where,
except for small pockets, the only models which do not violate $b
\rightarrow s \gamma$ are beyond the sensitivities of all present and
next generation direct experiments. One pocket in which this
constraint is not violated, near 60-100 GeV and $10^{-8}$ pb, are
those mSUGRA points which are gaugino mediated. Such features make
scenarios such as gaugino mediation and AMSB, which do not generally
violate this constraint, appear very attractive. 

\subsection{Comparison of Different Supersymmetry Breaking Scenarios}

\begin{table}[tbp] 
\begin{center}
\begin{tabular}{|c|c|c|c|c|c|c|c|}
\hline
&  & & & & & \\
Model & Direct &
 $\gamma$-G.C. & $\gamma$-G.C. & $\nu$-Sun & $e^+$ & $p^+$ \\
 & Searches & Cont.&  Line &  & & \\
\hline
\hline
mSUGRA & Good & Fair & Difficult & Difficult & Less Likely & OK \\
\hline
AMSB & Difficult & Good & Fair & Difficult & Likely & OK\\
\hline
Mich A, B & Fair & Difficult & Difficult & Difficult & Less Likely & OK \\
\hline
Mich C & Fair & Fair & Difficult & Difficult & Difficult & OK \\
\hline
Mich D, E & Difficult & Good & Difficult & Difficult & Likely & Fine Tuned\\
\hline
Mich F, G & Difficult & Difficult & Difficult & Difficult & Less Likley & OK\\
\hline
Gaugino & Fair & Difficult & Difficult & Difficult & Less Likely & OK \\
Mediation & & & & & & \\
\hline
SPS 1a & Good & Fair & Difficult & Difficult & Less Likely & OK\\
\hline
SPS 1b & Difficult & Difficult & Difficult & Difficult & Less Likely & OK\\
\hline
SPS 2    & Good & Fair & Difficult & Good & Likely & OK\\
Focus Point & & & & & &\\
\hline
SPS 3 & Good & Difficult & Difficult & Difficult & Less Likely & OK\\
\hline
SPS 4 & Good & Fair & Difficult & Difficult & Less Likely & OK\\
\hline
SPS 5 & Difficult & Difficult & Difficult & Difficult & Less Likely & OK\\
\hline
SPS 9 & Difficult & Good & Fair & Difficult & Less Likely & OK\\
\end{tabular}
\caption{A comparison of the detection potential and consistency of different SUSY
breaking models. For direct searches, continuum gamma-rays from the
galactic center, line emission gamma-rays from the galatic center and
neutrinos from the sun, each breaking scenario is described as
``Good'', ``Fair'' or ``Difficult''. These assessments reflect the
likelihood of detection with a given method. For the positron column,
``Likely'', ``Less Likely'' and ``Unlikely'' reflect the likelihood of neutralino annihilation in
the galactic halo being capable of producing the observed positron excess. Finally, the anti-proton column
evaluates whether the predicted anti-proton flux is compatiable with the BESS
measurement, or if a fine tuned halo distribution is needed to
reconcile such a model.   
\label{model_comp} }
\end{center}
\end{table}

From our numerical results, we have seen that different supersymmetry
breaking scenarios generically give rise to different dark matter
signatures since they predict different LSP properties as well as
different spectra for the other superpartners. 
We qualitatively summarize these differences in
Table~\ref{model_comp}. 

In Table~\ref{model_comp}, we list the mSUGRA and AMSB models first, as they are
the best studied scenarios. We then grouped the michigan benchmarks
into four groups based on the identity of the LSP.
The gaugino mediation scenario and the Snowmass benchmark points are
also listed. Although these points are allready covered in the scan of mSUGRA and AMSB, we find it
illustrative to study these benchmarks seperately as well.

%They should be already covered in the parameter space of
%mSUGRA. For example, the gaugino mediation scenario we look at are
%very close to the so-called no scale scenario of supersymmetry
%breaking. However, we list them independently as they represent
%distinct and interesting corners of the mSUGRA parameter space. 

We then assign qualitative measures such as ``good'', ``difficult'',
etc., to each supersymmetry breaking scenario regarding their
potential to produce observable signals in a particular
experiment. For the positrons column, we do not attempt to assess the liklihood of observing such a signal,
but rather estimate the plausibility that the observed positron excess could be generated from WIMP
annihilations for a given SUSY model. For the anti-protons column, we assess whether the SUSY model
requires fine tuning of the halo profile to remain consistent with the BESS data. 

Notice that mSUGRA is not
a point but rather a scenario with a sizable
parameter space. Therefore, the qualitative comments assigned to mSUGRA
should be understood as a general assessment applied to much of
the parameter space. Different corners of mSUGRA parameter space could
have quite different dark matter phenomenology. The same is true for our assessment of AMSB. Notice that
these comments are necessarily
qualitative and are not objective measures. For a more precise assessment, see the
figures we present in this paper rather than the table. 

A main lesson we can take away from Table~\ref{model_comp} is that  even if a
single experiment cannot provide a definitive answer to distinguish
between different supersymmetry breaking scenarios, a combination of
the results of all the experiments could be a powerful tool in this
respect. For example, several scenarios in this table could produce a
sizeable signal in future direct search
experiments. However, a observation (or non-observation) from the
neutrino signal from the Sun/Earth could favor (or disfavor) the focus
point region of the parameter space. Another example is the
combination of the positron signal with direct detection. As we
can see from the table, AMSB, Mich D,E, and the focus point region can
give rise to a sizeable positron flux. However, their direct detection rates differ quite
significantly. Notice that most of the
models we study will at least give rise to a potentially observable signal in
some experiment, and it will be possible to evaluate these scenarios based on the experimental results.

\section{Conclusions and Outlook}

\vspace{0.2in}

The large number of experiments designed to search for supersymmetric
dark matter particles are beginning to enter a very exciting
phase. Direct, indirect and collider experiments are each capable of a
great deal of progress in the near future. 

\begin{enumerate}
\item Direct Searches:

Direct searches are expected to improve their sensitivity by more
than a factor of $10^3$ with planned experiments such as ZEPLIN-4
\cite{zeplin4}, GENIUS \cite{genius}, or a 1 ton version of XENON \cite{xenon}. With such improvements, many scenarios, including much of mSUGRA,
will be probed by these
experiments. AMSB scenarios are more difficult to test by direct
experiments, although some models should be accessible. Gaugino mediated models may be testable in these planned experiments.

\item Neutrinos From Solar Capture: 

Indirect experiments which involve the capture of neutralinos in the
center of the Sun (or Earth) depend principally on the
neutralino-nucleon elastic scattering cross section for capture and,
therefore, have a tendency to be sensitive to the same models as
direct experiments. In fact, we have shown that for moderate or heavy
neutralinos ($\sim 150\,$ GeV or heavier), IceCube and ZEPLIN-4 have very
similar sensitivities. The focus point region of mSUGRA is, perhaps, the best prospect for this experimental method.

\item Gamma-Rays From The Galactic Center:

Rates for gamma-ray (as well as positron and anti-proton) experiments depend primarily on the neutralino
annihilation cross section. For these methods, the galactic halo
profile can be of enormous importance. Particularly, the sensitivity of gamma-ray
experiments, such as EGRET or GLAST, depend critically on the halo profile. Cuspy halo profiles often
provide observable gamma-ray rates in such
experiments, although profiles with lower density cores generally do not. Continuum emission of gamma-rays would be most easily observed in AMSB scenarios, although the prospects are still quite good for mSUGRA models. Gaugino mediated models are somewhat more difficult.  For line emission, however, gaugino mediated models provide better rates than other mSUGRA scenarios.

\item Positrons and Anti-Protons:

Cosmic positron and anti-proton
experiments do not depend
strongly on the size of the cusp, but rather on the smoothness or
clumpiness of dark matter in the local region, as well as the location of any nearby clumps. As was the case for gamma-ray rates, positron and anti-proton rates are largest in AMSB scenarios.

\item Flavor Constraints:

Flavor constraints, $b\rightarrow s \gamma$ in particular, can reveal
important information relevant to neutralino dark matter phenomenology. In
breaking scenarios without a solution to the flavor problem, SUGRA for
example, many models violate this constraint, especially
those which are most easily probed by direct experiments, or indirect experiments which use WIMP capture in the Sun. Scenarios which include a solution to
the flavor problem, such as AMSB, do not often have such features. 

\item Correlations and Comparisons:

By combining the results of multiple experimental techniques, characteristics of a particular supersymmetry breaking mechanism may be observed. For example, rates which depend primarily on the LSP annihilation cross section could be combined with rates which depend on the LSP-nucleon elastic scattering cross section. We find that by combining all the experimental data which can be
obtained in the foreseeable future, it may be possible to actually
make distinctions between a variety of SUSY breaking scenarios.

\item Other Future Prospects:

With the current operation of the Tevatron and future running of the
Large Hadron Collider (LHC), it is very likely that supersymmetry will
be discovered. With measurements of the supersymmetric spectrum (or
portions of it), the problem of predicting event rates for direct and
indirect dark matter experiments will likely become much more
simple. Additionally, astronomical measurements of the galactic halo
profile will shed a great deal of light on the prospects for indirect
detection. It appears that a great deal of information is beginning to
converge and an answer to these questions may not be far away. 

\end{enumerate}

%-----------------------------------------------------------------------
% THE ACKNOWLEDGEMENTS
%-----------------------------------------------------------------------
\section*{Acknowledgments}

We would like to thank Lars Bergstrom and Graham Kribs for insightful communications. We would also like to thank Tilman Plehn and Gordon Kane for useful comments on the draft. This research was supported in part by the U.S. Department of Energy
under Grants No. DE-FG02-95ER40896 and in part by the Wisconsin Alumni Research Foundation.

%-----------------------------------------------------------------------
%   REFERENCES
%-----------------------------------------------------------------------

%-----------------------------------------------------------------------
% FIGURES
%-----------------------------------------------------------------------
\newpage

% FIG. 1

% FIG. 2

%          

% FIG. 2b
%

%% FIG. 2c
%\begin{figure}
%\centering\leavevmode
%\epsfxsize=6in\epsffile{dirbm.ps}

%\caption[]{
%%
%Scalar (spin-independent) neutralino-nucleon elastic scattering cross section %verses neutralino mass. Each point corresponds to a theoretical prediction for% a specific Michigan benchmark model. Models A through G are shown. See the te%xt for more details.

%The experimental limits and sensitivities, shown as lines, from top to bottom %(on the right) are CDMS-March 2002 (solid) \cite{cdmsnow}, ZEPLIN 1-final 2002% (dashed) \cite{zeplin1}, Edelweiss-2000+2002 (dots) \cite{edelweiss}, CRESST %II-projected limit (s
%olid) \cite{crest}, CDMS-projected limit (dashed) \cite{cdms}, Edelweiss II-pr%ojected limit (dots) \cite{edelweiss2}, ZEPLIN 4-projection (solid) \cite{zepl%in4} and XENON-1 ton, projected limit (dashed) \cite{xenon}. Also shown, as a %solid contour, is the

% region which the DAMA collaboration claims discovery \cite{dama}. For a summa%ry of present limits and projections for all direct dark matter experiments, s%ee Ref.~\cite{alldirect}.
%
%\label{fig:twoc}
%
%}\end{figure}
%

% FIG. 3

%             

% FIG. 3b
%

%% FIG. 3c
%\begin{figure}
%\centering\leavevmode
%\epsfxsize=6in\epsffile{cenbm.ps}

%\caption[]{
%%
%The rate of continuum gamma-rays above 1 GeV per square meter, per year of exp%osure, from the galactic center, verses neutralino mass. A smooth NFW halo pro%file is assumed. If a Moore, {\it et. al.} model were considered, each point w%ould produce approximately a factor of 1000 more events.

% Each point corresponds to a theoretical prediction for a specific Michigan be%nchmark model. Models A through G are shown. See the text for more details.

%Also shown are limits from EGRET (solid) and predicted sensitivity for the fut%ure experiment GLAST (dashed) \cite{brenda}.
%
%
%\label{fig:threec}
%
%}\end{figure}
%

% FIG. 4
%

% FIG. 5
%

% FIG. 6
%

% FIG. 7
%

% FIG. 7b
%

%% FIG. 7c
%\begin{figure}
%\centering\leavevmode
%\epsfxsize=6in\epsffile{sunbm.ps}

%\caption[]{
%%
%The rate of muons (from charged current muon neutrino interactions) above 50 G%eV per square kilometer, per year, from neutralino annihilation in the sun, ve%rses neutralino mass. Each point corresponds to a theoretical prediction for a% specific Michigan be

%nchmark model. Models A through G are shown.
%See the text for more details. Also shown is the predicted sensitivity for the% next generation neutrino telescope IceCube (dashed) \cite{edsjolimits}.
%
%
%\label{fig:sevenc}
%
%}\end{figure}
%

% FIG. 8
%

% FIG. 11
%

% FIG. 13
%

% FIG. 9
%

% FIG. 10
%
%-----------------------------------------------------------------------
%   END DOCUMENT
%-----------------------------------------------------------------------
\end{document}